\documentclass[prb,twocolumn,showpacs,floatfix]{revtex4}

\usepackage{amsmath}
\usepackage{amssymb}
\usepackage{graphicx}
\usepackage[english]{babel}
\usepackage{blindtext}
\usepackage[utf8]{inputenc}

\usepackage{subfigure}

\newcommand{\CaO}{CaO}
\newcommand{\AlTwoOThree}{Al$_{\mathrm{2}}$O$_{\mathrm{3}}$}

\begin{document}


\title{Scattering of infrared light by dielectric core-shell particles}


\author{E. Thiessen, F. X. Bronold, R. L. Heinisch, and H. Fehske}
\affiliation{Institut für Physik, Ernst-Moritz-Arndt-Universität Greifswald, 17487 Greifswald, Germany}


\date{\today}

\begin{abstract}
We study the scattering of infrared light by small dielectric core-shell particles taking an 
\AlTwoOThree\ sphere with a \CaO\ core as an example. The extinction efficiency of such
a particle shows two intense series of resonances attached, respectively, to in-phase and 
out-of-phase multipolar polarization-induced surface charges build-up, respectively, at the 
core-shell and the shell-vacuum interface. Both series, the character of the 
former may be labelled bonding and the character of the latter antibonding, give rise to 
anomalous scattering. For a given particle radius and filling factor the Poynting vector field 
shows therefore around two wave numbers the complex topology of this type of light scattering. 
Inside the particle the topology depends on the character of the resonance. 
The dissipation of energy inside the particle also reflects the core-shell structure. It depends 
on the resonance and shows strong spatial variations.
\end{abstract} 


\maketitle

\section{Introduction}

Nanoparticles with a core-shell structure are of interest for various fields of applied 
research. They are routinely fabricated~\cite{HLL12,OA98} and have applications
ranging from nanophotonics~\cite{FZ14,P04} to medical treatment.~\cite{LL05} In the field 
of plasmonics,~\cite{O06} for instance, recent attention has been paid to nanoparticles 
having a metallic core and a dielectric shell~\cite{LMN12a,LMN12b,ZJB12} or vice 
versa.~\cite{FZ14,Stockman08} The reason is the added tunability of the optical response 
due to the geometry and composition of the particle. One of the most remarkable properties 
of coated nanoparticles is that they may induce transparency in a certain range of wave 
numbers due to the interplay of the dipole radiations of the core and the shell.~\cite{AE05} 

We are interested in the scattering of light by particles with a dielectric core and a dielectric 
coat (shell) where both materials show a strong transverse optical phonon resonance giving rise 
to anomalous scattering in the infrared occurring for wave numbers $\lambda^{-1}$ where the complex 
dielectric function $\varepsilon=\varepsilon^\prime+i\varepsilon^{\prime\prime}$ 
has $\varepsilon^\prime<0$ and $\varepsilon^{\prime\prime}\ll 1$.~\cite{TL06,Tribelsky11} 
In our previous work~\cite{THB14} we analyzed the extinction efficiency of this type of core-shell 
particles with an eye on using it in a low-temperature plasma as an electric probe with an optical 
read-out. The idea, originally put forward for homogeneous dielectric particles,~\cite{HBF12,HBF12109} 
is to utilize the blue-shift of the anomalous dipole resonance due to the surplus electrons collected from 
the plasma as a diagnostics from which the charge of the particle and thus the floating potential at the 
particle's position in the plasma can be determined. Whereas for homogeneous particles the charge-induced 
shift is most probably too small to be of practical importance, core-shell particles show a much 
larger blue shift. In particular the position of what we called shell resonance is very charge sensitive. 
Its blue-shift should be measurable for particle radii up to $0.5~\mu$m by infrared attenuation spectroscopy 
as it is typically used in plasma diagnostics.~\cite{RLR06} We also proposed to use dielectric
particles scattering infrared light in the anomalous regime as grains in dusty plasmas and to replace 
traditional force-balance techniques~\cite{CJ11,FPU04} of determining the charge of the grains by an 
optical technique. 

The dielectric core-shell particles suggested for an optical charge measurement contain a core with 
negative and a shell with positive electron affinity. This particular choice of electron affinities
results in a potential well which localizes the surplus electrons in the shell of the particle. 
Compared to an homogeneous particle, where the surplus electrons are spread out over the whole 
particle, the volume charge density of surplus electrons is thus enhanced. It is this localization effect
which already yields larger charge-induced shifts for the anomalous resonances of the core. More important
however is that the core-shell structure leads also to an additional resonance not present 
in an homogeneous particle.~\cite{Ruppin75,Uberoi80} Within the hybridization model for the optical 
response of complex nanostructures~\cite{Prodan03,Preston11} the additional resonance can be understood 
as the antibonding split-off of the anomalous resonance of the shell due to its mixing with a 
cavity mode supported by the core. The bonding partner is located at smaller wave
numbers with the splitting between the bonding and the antibonding resonance controlled by the filling 
factor $f=b/a$ which is the core radius $b$ measured in units of the particle radius $a$. In our 
investigations of charged core-shell particles we found the position of the antibonding dipole resonance 
(which we called shell resonance) to be particularly charge-sensitive.

The purpose of the present work is to analyze the bonding and antibonding resonances showing up in
dielectric core-shell particles in more detail using a neutral \AlTwoOThree\ particle with a \CaO\
core as an example. In addition to the results for this particular physical particle we also show 
results for a dissipationless model particle where the scattering anomalies can be more clearly 
identified. We demonstrate that the bonding as well as the antibonding resonances scatter light in
the anomalous regime. This can be deduced from the inverse hierarchy of the partial extinction 
cross sections and the topology of the Poynting vector field.~\cite{TL06}
In contrast to the hybridization model~\cite{Prodan03,Preston11} we do 
not use the electrostatic approximation. Instead we solve the full Mie 
equations~\cite{Mie08,Stratton41,Kerker69,BH83} for the core-shell particle.~\cite{AK51} We can 
thus gain further insight into the hybridization scenario~\cite{Prodan03,Preston11} by investigating 
specifically in the dipole regime the energy flux inside and outside the particle by the methods 
previously used for homogeneous particles.~\cite{WLH04,SVL04,BFZ05,LT06,LMT12} From the spatial 
distribution of the dissipation we can moreover identify regions inside the particle where 
most of the energy is deposited which may be of technological interest.~\cite{LMT12} Dissipation 
depends on the character of the resonance--bonding or antibonding--and is for both cases highly 
inhomogeneous, varying by two orders of magnitude from one position to another. 

The remaining paper is organized as follows. In the next section we recall briefly the Mie theory of 
light scattering by core-shell particles relegating mathematical details necessary to fix our notation 
to an appendix. In section~\ref{Results} we present our results, discussing first the hierarchy of 
the two series of extinction resonances and then--in the dipole regime--the topology of the Poynting
vector field, the magnitude and orientation of the electric field, and the dissipation of 
energy inside the particle. Concluding remarks with a summary of our main findings are 
given in section~\ref{Conclusions}.

\section{Theoretical background} \label{theory}
\begin{figure}[t]
  \centering
  \includegraphics[width=0.6\linewidth]{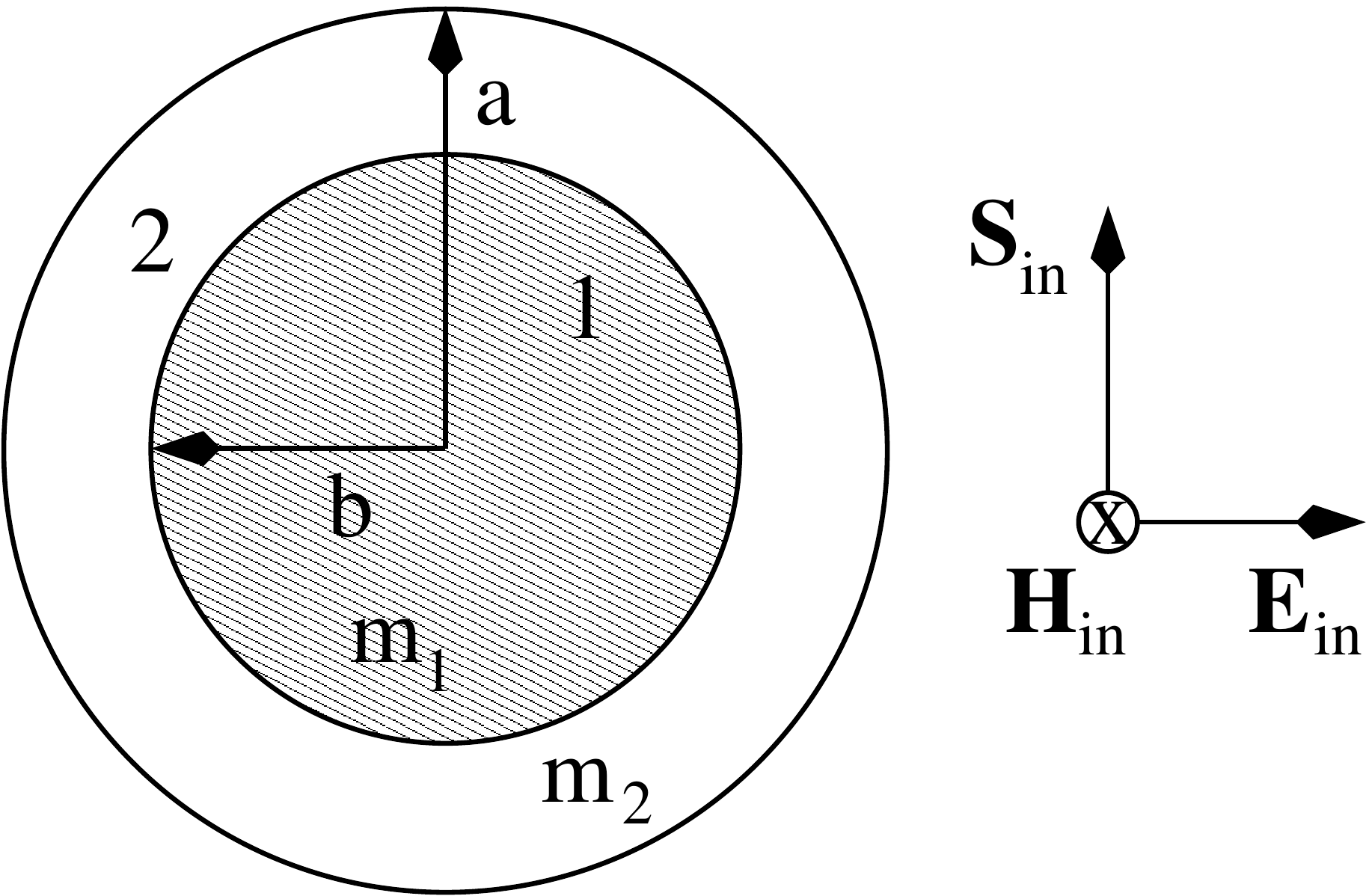}
  \caption{Illustration of the scattering geometry. The incident electromagnetic 
  plane wave is characterized by a Poynting vector ${\bf S}_{\rm in}$, an electric 
  field ${\bf E}_{\rm in}$, and a magnetic field ${\bf H}_{\rm in}$. It propagates 
  along the $z$-axis with an electric field polarized along the $x$-axis. The index
  of refraction in the core ($i=1$) and the shell ($i=2$) of the particle is 
  $m_{i}=\sqrt{\varepsilon_{i}}$, 
  where $\varepsilon_i$ is the complex dielectric function in region $i$. The total radius 
  of the particle is $a$ while the radius of the core is $b$. 
  }
  \label{ParticleDesign}
\end{figure}

In Fig.~\ref{ParticleDesign} we show the geometry of the scattering problem we consider. An
electromagnetic wave, linearly polarized in $x$-direction and propagating in $z$-direction,
hits a core-shell particle located at the center of the coordinate system. The particle is 
embedded in vacuum. It has a total radius $a$ and a core radius $b=fa$ where $f$ is the 
filling factor of the particle. In the limit $f\rightarrow 1$ the particle has thus no shell 
whereas for $f\rightarrow 0$ it has no core. The core ($i=1$) and the shell ($i=2$) of the 
particle are made out of different dielectric materials characterized by magnetic 
permeabilities $\mu_i=1$ and dielectric functions
$\varepsilon_i(\lambda^{-1})=\varepsilon_i^\prime(\lambda^{-1})+i\varepsilon_i^{\prime\prime}(\lambda^{-1})$
giving rise to refractive indices $m_i(\lambda^{-1})=\sqrt{\varepsilon_i(\lambda^{-1})}$, where 
$\lambda^{-1}=\nu/c$ is the wave number of the incident infrared radiation, $\nu$ is its frequency,
and $c$ is the speed of light. In the formulas given below and in the appendix we will use 
$\omega=2\pi\nu$ and $k=2\pi\lambda^{-1}$ instead of $\nu$ and $\lambda^{-1}$, the abbreviation
$k_i=k m_i$, and the size parameters $x=2\pi\lambda^{-1}b$ and $y=2\pi\lambda^{-1}a$. 

The theory of light scattering by an homogeneous spherical particle worked out originally 
by Mie~\cite{Mie08} has been extended to a core-shell particle by Aden and Kerker~\cite{AK51}
a long time ago. Since then the approach has been applied extensively~\cite{Ruppin75,Uberoi80} 
and it has also found its way into many textbooks.~\cite{Stratton41,Kerker69,BH83} 
Mathematically the Maxwell equations for the scattering problem shown in Fig.~\ref{ParticleDesign} 
are solved by expanding the electromagnetic fields in the three regions of 
interest--the outer space, the shell, and the core--in terms of spherical vector harmonics
and determining the expansion coefficients from the boundary conditions at the two interfaces.

Outside the particle there is the incident and the scattered wave. In terms of vector harmonics the 
fields of the incident wave can be written as~\cite{BH83}
\begin{align}
\vec{E}_{\rm in} &=\sum_{n=1}^\infty E_n \big( \vec{M}^{(1)}_{o1n} - i \vec{N}^{(1)}_{e1n} \big)~,\\
\vec{H}_{\rm in} &=-\frac{k c}{\omega\mu_m}\sum_{n=1}^\infty E_n \big( \vec{M}^{(1)}_{e1n}
            + i \vec{N}^{(1)}_{o1n} \big)
\end{align}
with expansion coefficients
\begin{align}
E_n=i^nE_0\frac{2n+1}{n(n+1)}~,
\end{align}
where $E_0$ is the strength of the incident electric field, eventually used as the unit for the 
field strength, $\mu_m=1$ is the vacuum permeability, and the superscript $(1)$ 
denotes that the radial dependence of the fields is given by the Bessel function $j_n$. The expansions for 
the fields of the scattered wave are  
\begin{align}
\vec{E}_{\rm s} &= \sum_{n=1}^\infty E_n \big( ia_n\vec{N}^{(3)}_{e1n} - b_n \vec{M}^{(3)}_{o1n} \big)~,\label{Es}\\
\vec{H}_{\rm s} &= \frac{k c}{\omega\mu_m}\sum_{n=1}^\infty E_n \big( ib_n\vec{N}^{(3)}_{o1n} \label{Hs}
+ a_n \vec{M}^{(3)}_{e1n} \big)~
\end{align}
with a superscript (3) indicating that the Bessel function $h_n$ describes now the radial dependence
of the fields. 

Inside the particle the fields in the shell and the core have to be distinguished. Inside the shell,
\begin{align}
\vec{E}_2 &= \sum_{n=1}^\infty E_n \big( f_n\vec{M}^{(1)}_{o1n} - ig_n \vec{N}^{(1)}_{e1n} \\
                                        &+ v_n\vec{M}^{(2)}_{o1n} - iw_n \vec{N}^{(2)}_{e1n} \big)~,\label{Eshell}\\
\vec{H}_2 &= -\frac{k_2 c}{\omega\mu_2}
          \sum_{n=1}^\infty E_n \big( g_n \vec{M}^{(1)}_{e1n} + if_n \vec{N}^{(1)}_{o1n} \\
            &+ w_n \vec{M}^{(2)}_{e1n} + iv_n \vec{N}^{(2)}_{o1n} \big)~, \label{Hshell}
\end{align}
while inside the core  
\begin{align}
\vec{E}_1 &= \sum_{n=1}^\infty E_n\big( c_n\vec{M}_{o1n}^{(1)} - id_n\vec{N}_{e1n}^{(1)} \big)~, 
\label{Ecore}\\
\vec{H}_1 &= -\frac{k_1 c}{\omega\mu_1}\sum_{n=1}^\infty E_n
\big(d_n\vec{M}_{e1n}^{(1)} + ic_n\vec{N}_{o1n}^{(1)} \big)~
\label{Hcore}
\end{align}
with the superscripts $(1)$ and $(2)$ indicating, respectively, to use the Bessel function $j_n$ and 
$y_n$ for the radial dependence of the fields. The expansion coefficients $a_n, b_n, c_n, d_n, f_n, 
g_n, v_n$, and $w_n$ have to be determined from the boundary conditions at the core-shell and the 
shell-vacuum interface giving rise to a system of eight equations for eight unknowns.~\cite{BH83}
Once the expansion coefficients are known quantities of physical interest can be computed. In the following 
we focus on the extinction efficiency, the Poynting vector field, and the dissipation of energy inside
the particle. 

Of central importance for the characterization of the scattering process is the extinction efficiency, 
that is, the scattered and absorbed energy per second divided by the incident energy flux per unit area. 
It is given by~\cite{BH83}
\begin{align}
Q_t=\sum_{n=1}^\infty Q^{(n)}_t~
\label{Qt}
\end{align}
with 
\begin{align}
Q^{(n)}_t=\frac{2}{y^2}{\rm Re} \left[(2n+1)(a_n+b_n)\right]
\label{Q_n}
\end{align}
the partial extinction efficiency for the multipole resonance of order $n$ where $n=1$ denotes 
the dipole resonance, $n=2$ the quadrupole resonance, and so on. 

Whereas $Q_t$ is of relevance only in the outer space the time-averaged Poynting vector field 
is of interest in the whole space. It can be obtained for the outer and the inner regions by 
inserting the expansions for the fields given above into the standard expression~\cite{Stratton41}
\begin{align}
\vec{S}=\frac{1}{2}\frac{c}{4\pi} \mathrm{Re}\left[\vec{E} \times \vec{H}^*\right]~.
\label{Poynting}
\end{align}
Likewise the dissipation of energy inside the particle can be obtained from~\cite{LL84}
\begin{align}
W=\frac{1}{2}\frac{\omega}{4\pi}\varepsilon'' \big|\vec{E}\big|^2~
\label{eq:dissip}
\end{align}
using for the electric field and the imaginary part of the dielectric function the values 
applicable to the region under consideration.

Outside the particle only the scattered and incident fields, $\vec{E}_{\rm s,in}$ and $\vec{H}_{\rm s,in}$,
are present. For the calculation of the extinction efficiency and the outside Poynting vector field the 
coefficients $a_n$ and $b_n$ are thus sufficient. In order to obtain however the dissipation and the
Poynting vector field inside the particle the fields $\vec{E}_{1,2}$ and $\vec{H}_{1,2}$ and hence the coefficients
$c_n, d_n, f_n, g_n, v_n$, and $w_n$ are also required. In the notation of Bohren and Huffman~\cite{BH83}
we list therefore all eight coefficients in the appendix.

\begin{table}[b]
\begin{center}
  \begin{tabular}{c|c|c|c}
    model  & $\lambda^{-1}_{\rm TO}$ (${\rm cm}^{-1}$) & ~~$\varepsilon_0$~~ & ~~$\varepsilon_\infty$~~ \\\hline
    core   & 300 & 3  & 2 \\
    shell  & 600 & 20 & 2 \\
  \end{tabular}
  \caption{Parameters for the model particle.}
  \label{Parameters}
\end{center}
\end{table}
%


\section{Results}
\label{Results}

We discuss the scattering of infrared light by neutral dielectric core-shell particles with filling 
factor $f=0.7$ for two particular cases: (i) a physical \CaO/\AlTwoOThree\ particle and (ii) an 
idealized model particle. For the former we take realistic bulk dielectric functions for the CaO 
core~\cite{Hofmeister03} and the \AlTwoOThree\ shell~\cite{Palik85,Barker63} to be found in our 
previous work~\cite{THB14} and plotted at the bottom of Fig.~\ref{shellres} while for the model 
particle the dielectric functions of the core and the shell are given by
\begin{align}
\varepsilon^\prime = \varepsilon_{\infty} + \lambda^{-2}_{\rm TO}\frac{\varepsilon_0 - \varepsilon_{\infty}}
                     {\lambda^{-2}_{\rm TO} - \lambda^{-2}}~,~~\varepsilon^{\prime\prime} = 0
\label{EpsModel}
\end{align}
with $\lambda^{-1}=\nu/c$ the wave number of the electromagnetic wave, $\lambda^{-1}_{\rm TO}$ the wave 
number of the transverse optical phonon and $\varepsilon_0$ and $\varepsilon_\infty$ the dielectric 
constants at low and high frequency, respectively. The values taken for the core and the shell are listed 
in Table~\ref{Parameters}. 

In contrast to the dielectric function of the physical \CaO/\AlTwoOThree\ particle the dielectric 
function of the model particle is real. The model particle has thus no damping. Since by definition
anomalous resonances occur for wave numbers for which $\varepsilon^\prime<0$ 
and $\varepsilon^{\prime\prime}\ll 1 $, that is, in situations where dissipative losses are much 
smaller than radiative losses,~\cite{TL06} the model particle displays the properties of anomalous 
resonances more clearly than the \CaO/\AlTwoOThree\ particle which of course has dissipation. In 
particular the disentangling of the complex topology of the energy flux is at first easier for the 
model particle.

\subsection{Hierarchy of anomalous resonances} 
\label{Hierarchy}

\begin{figure}[t]
 \includegraphics[width=\linewidth]{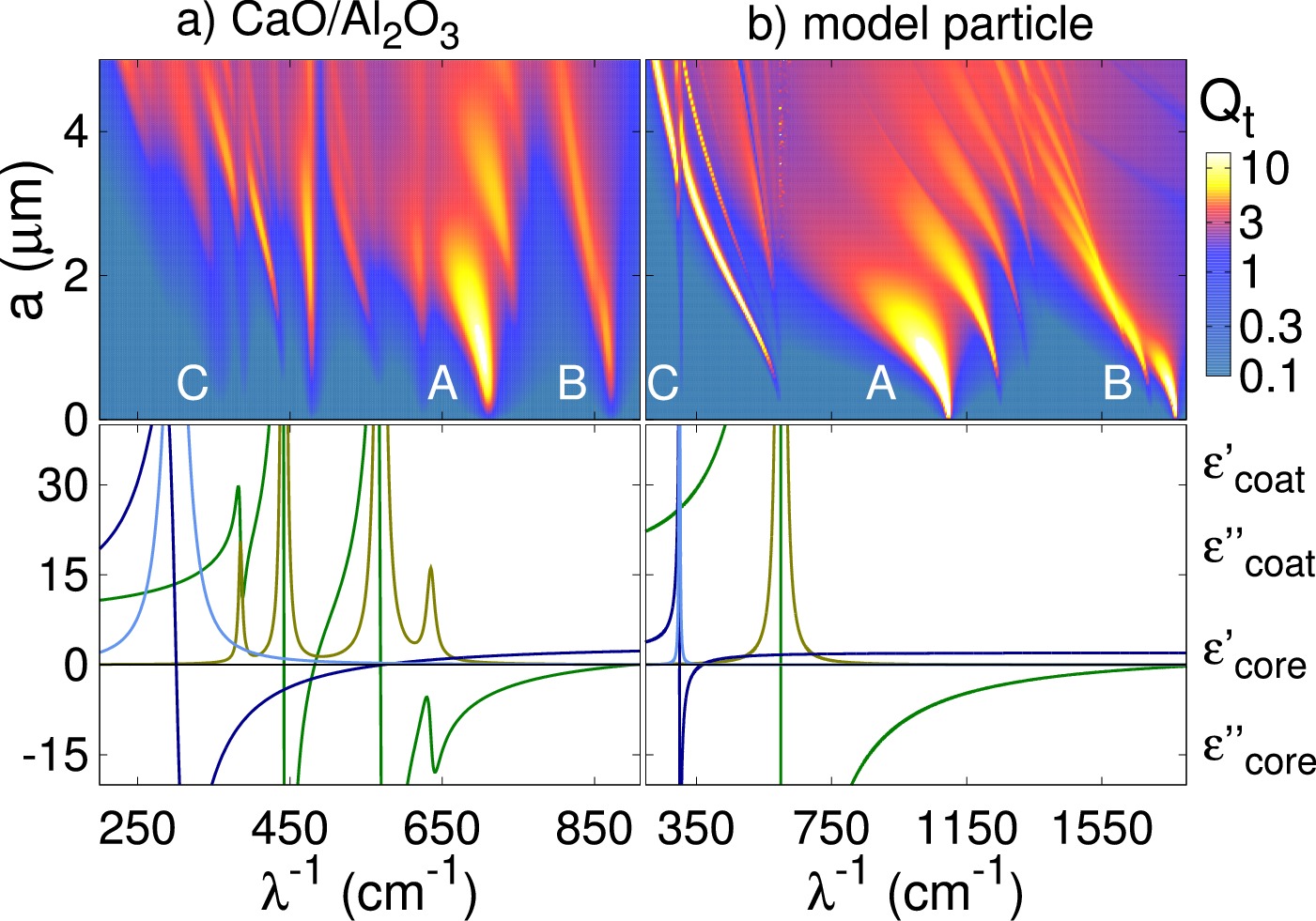}%
 \caption{\label{shellres} (Color online) Top: Extinction efficiency $Q_t$ as a function of the wave number 
 $\lambda^{-1}$  and the particle radius $a$ for filling factor $f=0.7$. The anomalous resonance of the core 
 is indicated by C while the two resonances appearing in the surface phonon regime of the shell are denoted 
 by A and B. Bottom: Real and imaginary parts of the dielectric functions used for the
 \CaO/\AlTwoOThree\ particle and the model particle, respectively.}
\end{figure}

We start with the extinction spectrum in the far-field of the particle. In Fig.~\ref{shellres} 
the extinction efficiencies $Q_t$ of the \CaO/\AlTwoOThree\ and the model particle are shown as a 
function of wave number $\lambda^{-1}$ and particle radius $a$ together with the dielectric functions
of the particles. Here and in the figures below the filling factor $f$ is always $0.7$. In the 
dipole regime, for very small radii, three resonances A, B, and C can be identified for both 
kinds of particles. Resonance C, which we will not discuss in detail, is the anomalous dipole resonance
of the core while resonance B is the one we proposed to use as a charge diagnostics.~\cite{THB14}
The resonances A and B are of particular interest because they belong to a series of resonances 
lying in the surface phonon regime of the shell, that is, in the spectral region between 
$\lambda^{-1}_{\rm TO,2}$ and $\lambda^{-1}_{\rm LO,2}$, where
$\lambda^{-1}_{\rm TO (LO),2}$ is the wave number of the transverse (longitudinal) optical phonon 
of the shell. The goal of our work is to investigate the physical properties of these two 
series. 

\begin{figure}[t]
 \includegraphics[width=\linewidth]{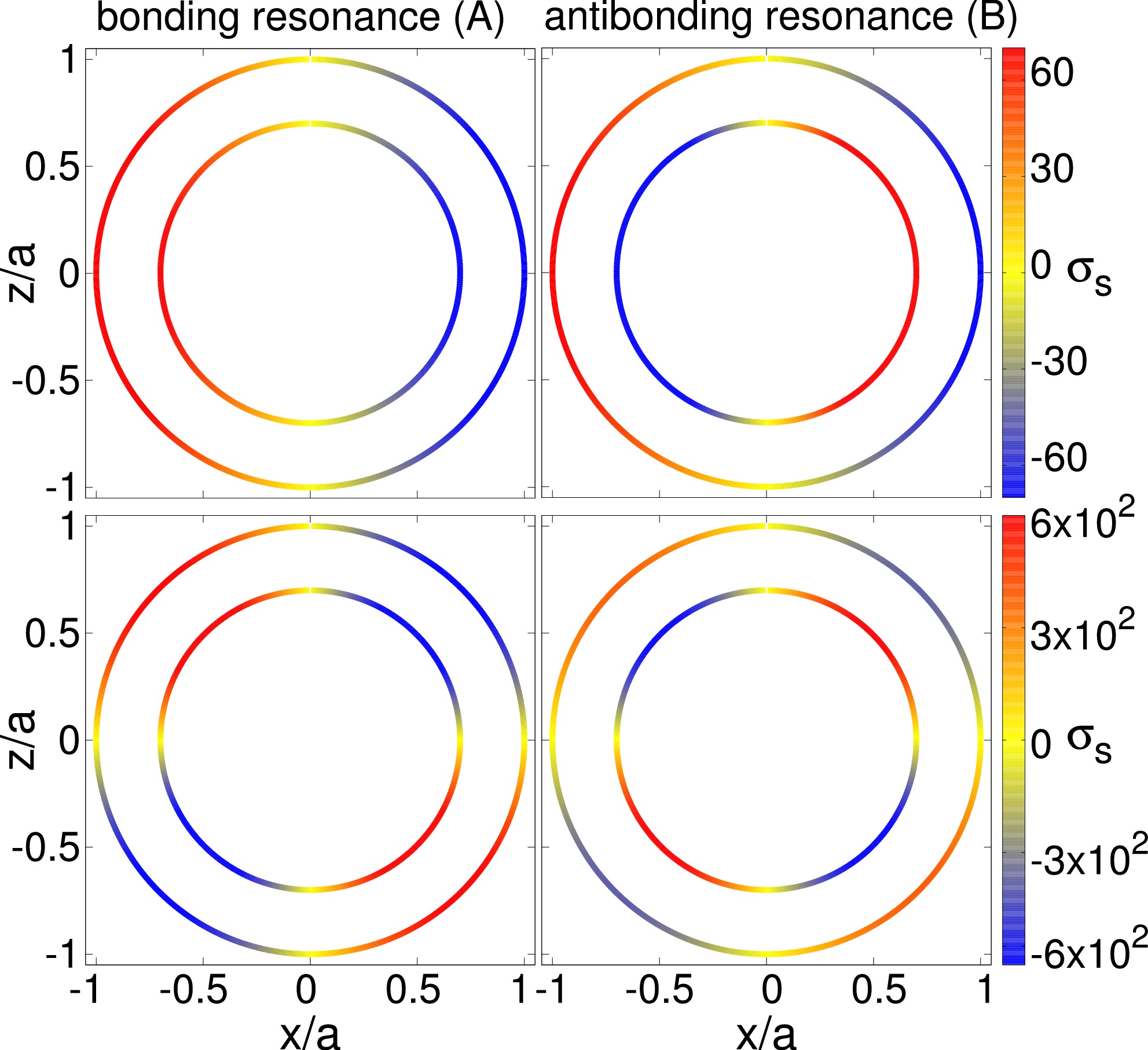}%
 \caption{\label{SurfaceMode} (Color online) Polarization-induced surface charges build-up at the  
  core-shell and the shell-vacuum interface of the model particle to satisfy the boundary conditions 
  of the Maxwell equations for the dipole (upper two panels) and the quadrupole resonances (lower 
  two panels). For the bonding resonances the multipolar surface charges of the two interfaces 
  are in-phase whereas for the antibonding resonances they are out-of-phase. The wave numbers
  are always slightly above the resonance: $\lambda^{-1}=1095.669\,{\rm cm}^{-1}$ (bonding 
  dipole), $\lambda^{-1}=1767.293\,{\rm cm}^{-1}$ (antibonding dipole), 
  $\lambda^{-1}=1241.95\,{\rm cm}^{-1}$ (bonding quadrupole), 
  and $\lambda^{-1}=1682.42\,{\rm cm}^{-1}$ (antibonding quadrupole). 
  The particle radius $a=0.1\,\mu m$ for the dipole and $a=0.6\,\mu m$ for the quadrupole 
  resonances with filling factor $f=0.7$ in both cases.
  }
\end{figure}

From the viewpoint of the hybridization model~\cite{Prodan03,Preston11} the two series arise 
from the mixing of a surface mode of a hypothetical particle with radius $a$ made out of 
shell material with a surface mode of a hypothetical cavity of radius $b$ made out of 
core material and embedded into a junk of shell material leading in each multipole order 
to a bonding and an antibonding mode. The hybridization scenario is buried in the Mie equations 
but it can be made explicit by plotting the difference of the normal components of the polarization,
$\vec{P}=\varepsilon_0(\varepsilon-1)\vec{E}$, at the core-shell and the shell-vacuum 
interface; $\varepsilon_0=1$ is the dielectric function of the vacuum. As can be seen in 
Fig.~\ref{SurfaceMode} for the particular case of the dipole and quadrupole resonances of 
the model particle, the polarization-induced surface charges required to satisfy the boundary 
conditions of the Maxwell equations at the two interfaces are for the bonding resonances in-phase 
and for the antibonding resonances out-of-phase. The two dipole resonances, for instance, can thus 
be understood in terms of an inner and an outer dipole which are oriented either parallel or antiparallel. 
In the vicinity of these two resonances the electric field outside the particle is thus the effective 
field of two parallel or antiparallel dipoles. Inside the particle the field is more 
complex (see below). Similarly the bonding and antibonding quadrupole resonances originate from 
quadrupole surface polarization charges at the two interfaces which are either in- or out-of-phase.

The parameters of the dielectric functions of the model particle have been chosen in such a 
way that the series of antibonding resonances can be also clearly identified. 
Both the bonding and antibonding series stop at the third multipole order,
that is, at the octupole, as it is also the case for the 
\CaO/\AlTwoOThree\ particle. But it is conceivable to have material combinations
and perhaps filling factors for which the series of sharp resonances continues to higher order. 
The question then arises whether the two series of resonances scatter light anomalously. 
In particular, it would be interesting to know whether the sharp resonances
signal the resonant excitation of a particular surface mode of the shell
and whether the hierarchies of extinction cross sections to which the two series give rise to are  
inverted as it would be the case for homogeneous particles made out of either the shell or the core 
material. 

\begin{figure}[t]
 \includegraphics[width=\linewidth]{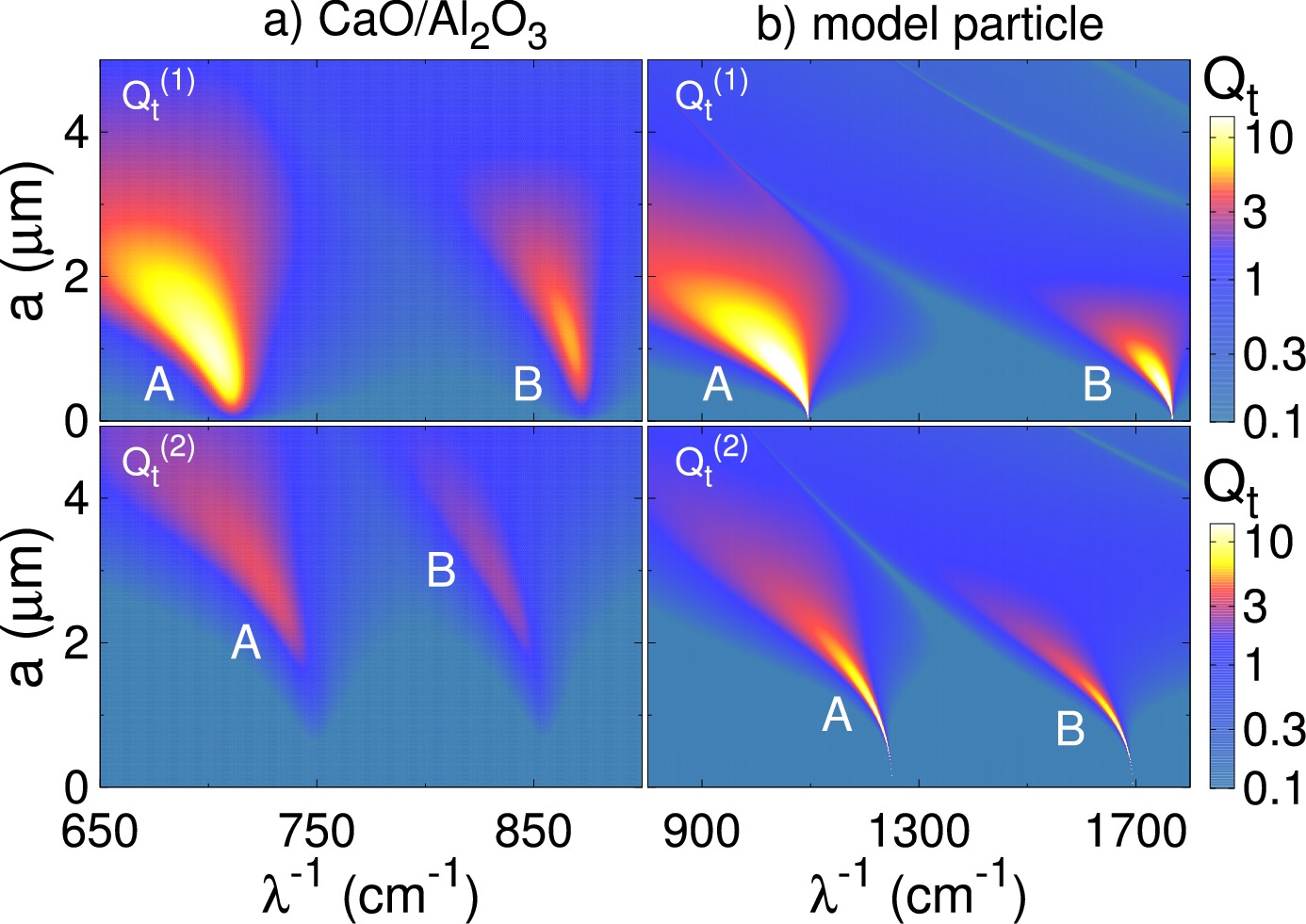}%
 \caption{\label{einzelneKoeff} (Color online) Partial extinction efficiency $Q_t^{(n)}$ in the vicinity 
  of the bonding (A) and antibonding resonance (B) as a function of the wave number $\lambda^{-1}$ and the 
  particle radius $a$. The superscripts $(1)$ and $(2)$ indicate the dipole and the quadrupole mode. Higher
  multipole modes appearing at larger radii are weaker and cannot be seen on the scale used for the
  plot. The filling factor $f=0.7$.}
\end{figure}

For an homogeneous particle Tribelsky and Luk'yanchuk~\cite{TL06} have shown that if the size parameter 
is at the resonance wave number of the order one or smaller the main contribution to this resonance 
comes from one resonantly excited surface mode. This selectivity remains for a core-shell particle, as we will
now demonstrate. For that purpose we plot in Fig.~\ref{einzelneKoeff} for the physical and the model particle 
the extinction efficiencies due to the dipole mode $Q_t^{(1)}$ and the quadrupole mode $Q_t^{(2)}$ separately. 
As can be seen each mode has its own distinct set of particle radii and wave numbers at which it is resonantly 
excited. Off-resonance the partial multipole extinction efficiencies provide only an homogeneous background
to the overall extinction efficiency. Thus, it is possible to relate to the lowest resonance in Fig.~\ref{shellres} 
an excited dipole mode and to the second lowest resonance an excited quadrupole mode. For a given particle 
radius a core-shell particle gives thus rise to two wave numbers at which a dipole mode or a quadrupole 
model can be excited. This is in contrast to an homogeneous particle where only one wave number exists.

Having identified the multipole order of the resonances we can now turn to the extinction 
cross section $\sigma_t=\pi a^2 Q_t$ and study its hierarchy. For an homogeneous particle $\sigma_t$ shows 
in the anomalous regime an inverse hierarchy, that is, for a given particle radius the cross section due 
to a higher multipole is larger than the cross section due to a lower multipole.~\cite{TL06,Tribelsky11} 

\begin{figure}[t]
\includegraphics[width=\linewidth]{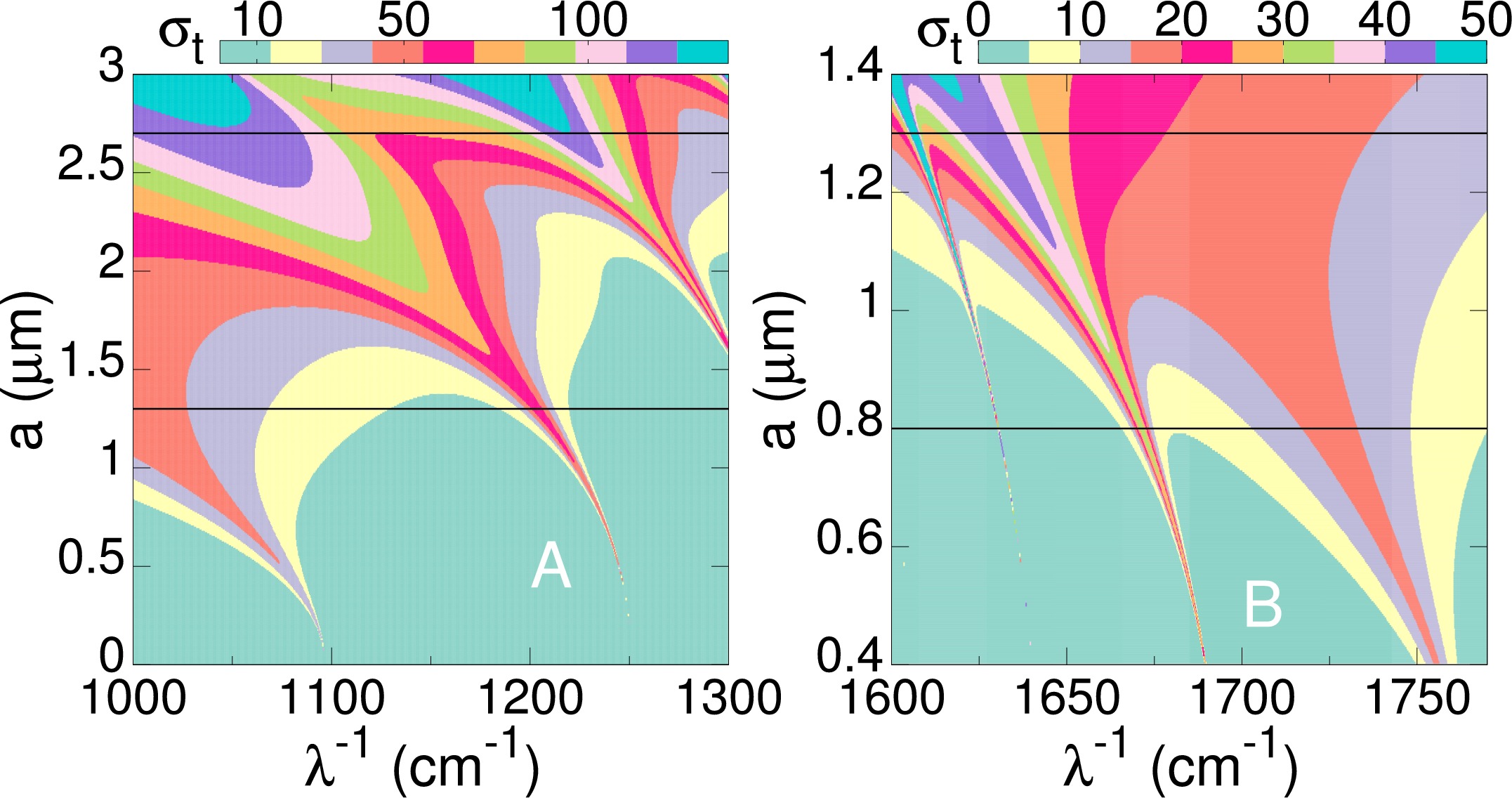}%
\caption{\label{crossSec} (Color online) Inverse hierarchy of the extinction cross section $\sigma_t$ in units 
of $\pi a^2$ for the model particle in the vicinity of the bonding (A) and the antibonding resonance (B). 
Horizontal lines indicate radii where the local maximum of the cross section shifted from a lower to a 
higher mode. The filling factor $f=0.7$.}
\end{figure}

In Fig.~\ref{crossSec} we see for the model particle that the series of bonding and antibonding resonances 
arising in core-shell particles display the same behavior. At the bonding resonance (left panel) for particle 
radii smaller than $a=1\,\mu$m the cross section $\sigma_t$ of the dipole resonance at 
$\lambda^{-1}\approx 1050\,{\rm cm}^{-1}$ 
is dominant, at $a=1.3\,\mu$m on the other hand, the maximum of the cross section shifted to the quadrupole 
resonance at $\lambda^{-1}\approx 1200\,{\rm cm}^{-1}$. Yet another shift occurred for $a=2.7\,\mu$m, where the 
maximum of the cross section is now at the octupole resonance. This inverse hierarchy can also be seen for 
the antibonding resonance (right panel). At the particle radius $a=0.8\,\mu$m the cross section maximum has 
shifted from the dipole mode around $\lambda^{-1}\approx1740\,{\rm cm}^{-1}$ to the quadrupole mode around 
$\lambda^{-1}\approx 1680\,{\rm cm}^{-1}$. The next shift of the cross section maximum is indicated at $a=1.3\,\mu$m
where the octupole resonance is the most pronounced in the cross section. Since the inverse hierarchy is a direct 
consequence of the series of resonances in the extinction efficiency,~\cite{TL06,Tribelsky11} the real 
\CaO/\AlTwoOThree\ particle shows the same inverted hierarchy of resonances in the cross section (not shown 
in Fig.~\ref{crossSec}). 

Both the series of bonding and the series of antibonding resonances fulfill therefore the criteria 
of anomalous light scattering as spelled out by Tribelsky and Luk'yanchuk.~\cite{TL06,Tribelsky11} 
Each series consists of strong narrow resonances 
which are determined by a single resonantly excited bonding or antibonding mode and which moreover give 
rise to extinction cross sections increasing for fixed radius with the order of the multipole. The 
different physical origin of the two series, in-phase vs. out-of-phase polarization-induced surface charges
at the two interfaces, is reflected in the way the two series shift in wave number with increasing multipole 
order. Whereas the bonding resonances shift to higher wave numbers the antibonding resonances shift in the 
other direction. The antibonding quadrupole resonance is thus excited at lower wave numbers than the 
antibonding dipole resonance. 

\subsection{Topology of the Poynting vector field}

In the previous section we showed that the series of bonding and antibonding resonances have anomalous
properties. We now investigate in the vicinity of the two lowest resonances, the bonding and 
antibonding dipole resonances, the energy flux outside and inside the core-shell particle and 
compare it with the results Tribelsky and Luk'yanchuk obtained for an homogeneous particle.~\cite{TL06}
In particular, we consider a model particle with radius $a=0.1\,\mu$m and a \CaO/\AlTwoOThree\
particle with radius $a=0.4\,\mu$m keeping the filling factor in both cases fixed to $f=0.7$. As 
can be deduced from Fig.~\ref{einzelneKoeff} only the dipole resonances
are then excited. For the model particle they are, respectively, at 
$\lambda_{\rm b,1}^{-1}\approx 1095.669 \,{\rm cm}^{-1}$ 
and $\lambda_{\rm ab,1}^{-1}\approx 1767.293\,{\rm cm}^{-1}$, where the subscript stands for 
bonding/antibonding (b/ab) dipole ($n=1$) resonance, while for the \CaO/\AlTwoOThree\ 
particle $\lambda_{\rm b,1}^{-1}\approx 709.63\,{\rm cm}^{-1}$
and $\lambda_{\rm ab,1}^{-1}\approx 871.62\,{\rm cm}^{-1}$.

Tribelsky and Luk'yanchuk~\cite{TL06} investigated the Poynting flux in the vicinity of an anomalous 
resonance of an homogeneous particle and identified its singular points. Due to the large changes in 
the modulus and phase of the scattering coefficients in this spectral region the position and the 
occurrence of the singular points is very sensitive to little changes in the wave number of the 
incident light. A singular point is characterized by $|\vec{S}|=0$.~\cite{SVL04} For the scattering 
geometry shown in Fig.~\ref{ParticleDesign}, which is also the one used by Tribelsky and Luk'yanchuk,~\cite{TL06} 
the singular points are located in the $xz$-plane. Since in this plane $S_\phi(x,z)=0$ they are given 
by the intersection of the isoclines $S_r(x,z)=0$ and $S_\theta(x,z)=0$. In the dipole regime only the
coefficients $a_n$, $d_n$, $g_n$, and $w_n$ are resonant. Hence, the orientation of the electric and 
magnetic fields in space is in all three regions determined by the vector harmonics $\vec{N}^{(...)}_{e1n}$
and $\vec{M}^{(...)}_{e1n}$, respectively. The difference is in the radial dependence, signalled 
by the superscript $(...)$ which may stand for (1), (2), or (3) depending on the field under consideration. 

We also identify the type of the singular point which in vacuum, where $\nabla\cdot\vec{S}=0$, can 
be only a center, a saddle, a saddle-focus or a saddle-node.~\cite{WLH04} For that purpose 
we (i) carefully analyze the field lines around the singularity as each type of singularity 
shows a characteristic field distribution and (ii) we check the behavior of the magnetic 
field $H_y(x,z)=|H(x,z)| \mathrm{e}^{\mathrm{i}\phi(x,z)}$ as suggested by 
Schouten.~\cite{SVL04} At a center or a focus singularity the amplitude of the magnetic field 
vanishes $|H(x,z)|=0$, while a saddle or node point is characterized by $\nabla \phi(x,z)=0$.
Singular points due to anomalous resonances all lie in the near field since the characteristic 
spatial scales are smaller than the wave length of the radiation.~\cite{TL06} For instance, 
for the model particle, the largest distance of singular points due to the antibonding dipole 
resonance is about $r=1.5 \times \,10^{-4}\,$cm while the wavelength of the incident light is  
$\lambda=5.7\times\,10^{-4}\,$cm. Similarly, for the bonding dipole resonance, $r=2.5 \times \,10^{-4}\,$cm 
and $\lambda=9.13\times\,10^{-4}\,$cm. 

\begin{figure*}
 \includegraphics[width=\linewidth]{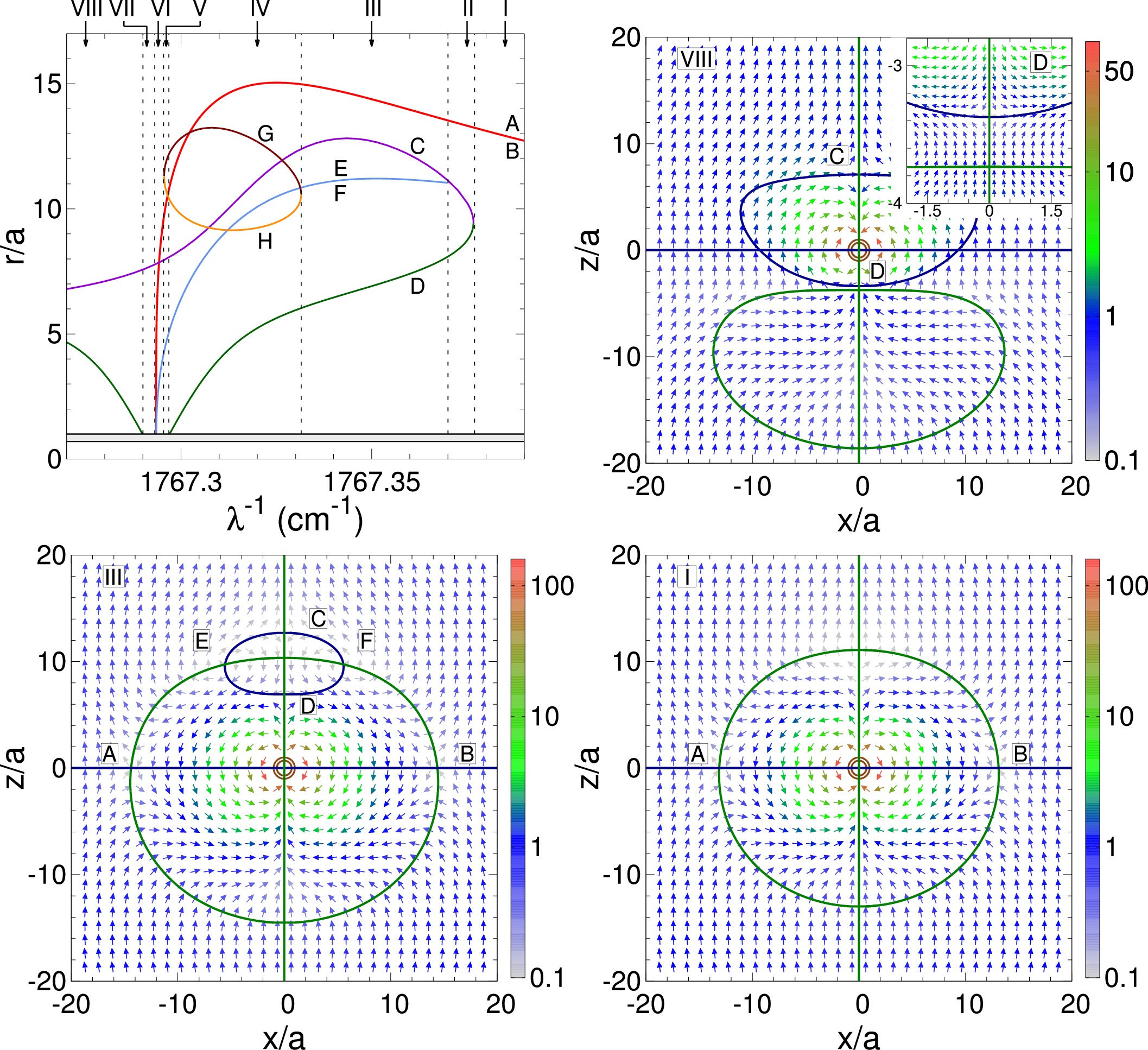}%
 \caption{\label{SRaussen} (Color online) Top left: Position of the outside singular points in the Poynting 
  field (Arabic letters) of the model particle with $f=0.7$ and $a=0.1\,\mu{\rm m}$ near the antibonding
  resonance at $\lambda^{-1}_{\rm ab,1}\approx 1767.293\,{\rm cm}^{-1}$ as a function of wave number.
  For the bonding resonance the results are similar. Spectral regions with a fixed topology are 
  indicated by Roman numbers and the shaded area denotes the shell. The other three panels show 
  the Poynting field for three fixed wave numbers. The color of the arrows gives the modulus of 
  the Poynting vector normalized to the flux of incident energy. The intersections of the isoclines 
  $S_\theta(r, \theta)=0$ (green) and $S_r(r, \theta)=0$ (blue) determine the singular points. 
  The shell and the core of the particle are indicated by brown circles.}
\end{figure*}
\subsection*{Topology outside the particle}

We start with the outside energy flux. For both the model and the \CaO/\AlTwoOThree\ 
particle only the dipole resonances are excited. Since outside the particles the topology of the energy 
flux in the vicinity of the bonding and antibonding (dipole) resonances turn out to be qualitatively 
the same, we discuss only results for the latter.

Figure~\ref{SRaussen} summarizes for the model particle the topology of the outside Poynting 
vector field in the vicinity of the antibonding dipole resonance ($a=0.1\mu$m, $f=0.7$, 
$\lambda^{-1}_{\rm ab,1}\approx 1767.293\,{\rm cm}^{-1}$). Note, the dramatic modifications
in the topology of the Poynting field due to very small changes in the wave 
number. The top left of the figure shows the position of the singular points, labelled by Arabic 
letters, with respect to the particle center. The Roman numerals on the top indicate regions 
with a fixed topology. The remaining three panels show the Poynting flux outside the particle, 
symbolized by the two brown circles in the center of the plots, for  
wave numbers representative for the indicated topological region. The arrows show the direction 
of the Poynting vector, while the colors correspond to the modulus of the Poynting flux normalized 
to the incident flux of light. In addition the $S_r=0$ isocline and $S_{\theta}=0$ isocline are 
indicated by blue and green lines. The singular points are given by the intersection of two 
differently colored isoclines. 

For wave numbers $\lambda^{-1}<\lambda_{\rm ab,1}^{-1}$, for instance, in region VIII, the Poynting 
flux in the vicinity of the particle shows a dipole radiation oriented opposite to the incident 
light which is propagating in $z$-direction. As an illustration of this behavior we plotted 
in the upper right panel of Fig.~\ref{SRaussen} the Poynting flux for 
$\lambda^{-1}=1767.28\,{\rm cm}^{-1}$. It contains two singular points, one in 
the upper and one in the lower half of the $xz$-plane. At the saddle point D in the lower half plane 
(replotted on a magnified scale in the inset) it can be nicely seen that the energy flux emitted 
by the particle due to the excited dipoles is oriented oppositely to the incident energy flux. The 
field lines emitted by the particle (positive $S_r$-component) and the 
field lines from the incoming wave (negative $S_r$-component) meet at this point and are then bend 
over to flow towards plus or minus $x$. In the upper half-plane, that is for $z\ge 0$, the Poynting 
flux either passes by or is reabsorbed by the particle. This can be seen at point C, which is the 
Bohren saddle point,~\cite{WLH04} where the field lines coming from $\pm x$ are going behind the 
particle (positive $S_r$-component) or towards it (negative $S_r$-component). The saddle points C 
and D reflect the strength of the energy flow of the resonant excitation relative to the strength 
of the energy flow of the incoming light. Thus, these points can disappear if the dipoles are not 
emitting enough energy, as it can happen if the damping inside the particle is too large.

On the high energy side of the resonance, for wave numbers $\lambda^{-1}>\lambda_{\rm ab,1}^{-1}$, 
the dipoles radiate energy parallel to the incident light. The Poynting flux thus gets more 
complicated. For example in the topological region III, at  $\lambda^{-1}=1767.35\,{\rm cm}^{-1}$, 
six singular points, four saddle points A, B, C, D, and two foci points E and F (in three 
dimensions these are saddle-focus points) arise, all located in the upper half of the 
$xz$-plane (see lower left panel of Fig.~\ref{SRaussen}). There are no singular 
points anymore for $z<0$. Since the excited dipoles changed simultaneously their directions
the $S_r$-component of the Poynting vector is always negative for $z<0$, that is, the energy 
flux along the $z$-axis is oriented towards the particle. One of the $S_r=0$ isoclines at which a 
change of sign in the $S_r$-component occurs is located above the particle at $z>0$ 
and so is now the saddle point D. At this point the field 
lines emitted by the particle and thus having a positive $S_r$-component 
and the field lines coming from the foci points E and F and heading towards the particle owing to 
the negative $S_r$-component meet and take their way towards plus and minus $x$. This is quite 
similar to the situation at point D in region VIII, since the foci points can be regarded as sources 
of energy in the $xz$-plane.~\cite{WLH04} At the Bohren saddle point C the field lines come together 
from the left and right focus and then they either go behind the particle or towards it, as they 
did in region VIII. At $z=0$ there are two saddle points A and B. To understand their functioning 
lets take a look along the $S_{\theta}=0$ isocline. For $z<0$ field lines which cross the 
$S_\theta=0$ isocline enter the particle while field lines crossing the $S_\theta=0$ isocline 
for $z>0$ leave the particle. Thus, at the singular points A and B field lines leaving or entering 
the particle come close together and form a saddle point.

Some wave numbers away from $\lambda^{-1}_{\rm ab,1}$, in region I, only the saddle points A 
and B remain as can be seen in the panel on the lower right of Fig.~\ref{SRaussen} where the
Poynting flux is plotted for $\lambda^{-1}=1767.38\,{\rm cm}^{-1}$. If we would plot separatrixes 
inside the circle defined by the $S_\theta=0$ isocline we would find Tribelsky ears.~\cite{WLH04}
For $z<0$ the Poynting flux has not changed qualitatively compared to region III. But for 
$z>0$ the $S_r=0$ isocline disappeared. As a result, there are no further singular points
other than A and B.

\begin{figure*}
 \includegraphics[width=0.9\linewidth]{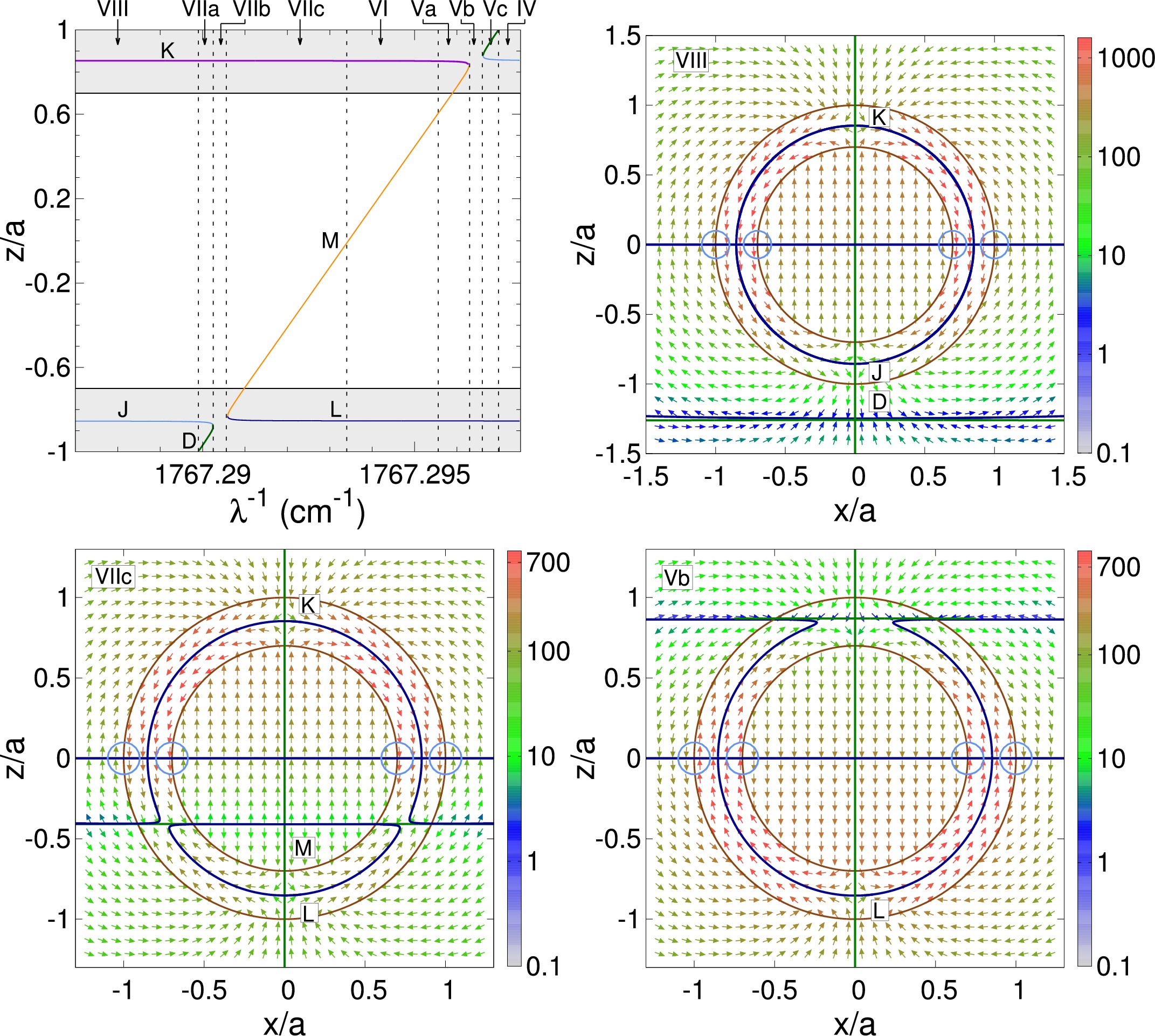}%
 \caption{\label{SRInside} (Color online) Top left: Position of the singular points (Arabic letters) occurring
 in the Poynting field inside the model particle ($f=0.7$ and $a=0.1\,\mu{\rm m}$) in the vicinity 
 of the antibonding dipole resonance at $\lambda^{-1}_{\rm ab,1}\approx 1767.293\,{\rm cm}^{-1}$. 
 The different topologies are indicated by Roman numbers and the shaded regions denote the shell. 
 The other three panels show the Poynting flux for three wave numbers, each one is 
 representative for a certain topology. The arrows show the direction of the Poynting flux at each 
 point, while the modulus of the Poynting vector normalized to the incident flux of energy is encoded 
 by the color of the arrows. The intersections of the isoclines $S_\theta(r, \theta)=0$ (green) and 
 $S_r(r, \theta)=0$ (blue) determine the singular points. The shell and the core are again indicated 
 by brown circles and the optical whirlpools are marked by light blue circles.}
\end{figure*}

A comparison of our results for the Poynting flux near the antibonding dipole resonance of a core-shell
particle with the results found by Tribelsky and Luk'yanchuk~\cite{TL06} near the anomalous resonance of 
a homogeneous particle shows that the topologies are the same. The same pattern arises also around the 
bonding dipole resonance. Hence, the changes in the anomalous topology of the Poynting flux 
appear for a core-shell particle around two wave numbers. By just looking at the energy distribution outside 
the particle one could not distinguish between a core-shell and an homogeneous particle. Although
in the first case two dipoles emit energy, which may have parallel or antiparallel 
orientation, depending on whether one is close to the bonding or close to the antibonding
dipole resonance, whereas in the second case the energy is emitted only by one dipole.  

\subsection*{Topology inside the particle}

The core-shell structure of the particle reveals itself in the energy flux inside the particle.
The reason is the relative orientation of the inner and the outer dipole. It is parallel for 
the bonding and antiparallel for the antibonding resonance. Below we will verify the orientation
of the dipoles explicitly by looking at the electric field lines. At present it suffices to 
recall the polarity of the surface polarization charges shown in Fig.~\ref{SurfaceMode} from which 
the orientation of the dipoles can be also deduced. Unlike the Poynting flux outside the
particle, which arises from the superposition of the incoming field with an effective overall 
dipole field of the particle, inside the core-shell particle the energy fluxes of two 
dipoles which are either parallel (bonding) or antiparallel (antibonding) mix. 

As explained at the beginning of this section in the dipole regime the singular points of the 
Poynting flux are all located in the $xz$-plane. The figures summarizing the results for the
Poynting flux inside the particle are thus constructed in analogy to Fig.~\ref{SRaussen}. We
will show the position of the singular points with respect to the particle center and for 
representative wave numbers the topology of the energy flux. The only difference is that the 
radial distance of the singular points from the center is now less than the particle radius $a$. 

\subsubsection{Antibonding dipole resonance}

The inside topology of the Poynting flux near the antibonding dipole resonance of the model 
core-shell particle is presented in Fig.~\ref{SRInside}. On the scale of this figure the 
localized surface electromagnetic modes characteristic for the type of light scattering 
we discuss in this paper can be clearly identified. They are associated with the optical 
whirlpools~\cite{BFZ05} in the Poynting flux which are indicated by light blue circles. The 
energy is rotating around whirlpools and thus $S_r=0$. However, since $S_\theta$ has a 
maximum~\cite{LT06} at these points whirlpools are not ordinary singularities. At the two 
interfaces the whirlpools are oriented in the 
opposite direction. It is instructive to trace the direction of the whirlpools to the 
orientation of the inner and the outer dipole. The polarity of the surface charges at the 
two interfaces as well as the electric field lines to be presented in Sec.~\ref{Efield} show 
that at the antibonding resonance the two dipoles are antiparallel. Hence, in the shell region 
the electric field lines belonging to the outer and the inner dipole are parallel. In the vicinity 
of the whirlpools this leads then to a rectifying Poynting flux inside the shell, with Poynting 
field lines separated by a $S_r=0$ isocline.

%
\begin{figure*}
 \includegraphics[width=\linewidth]{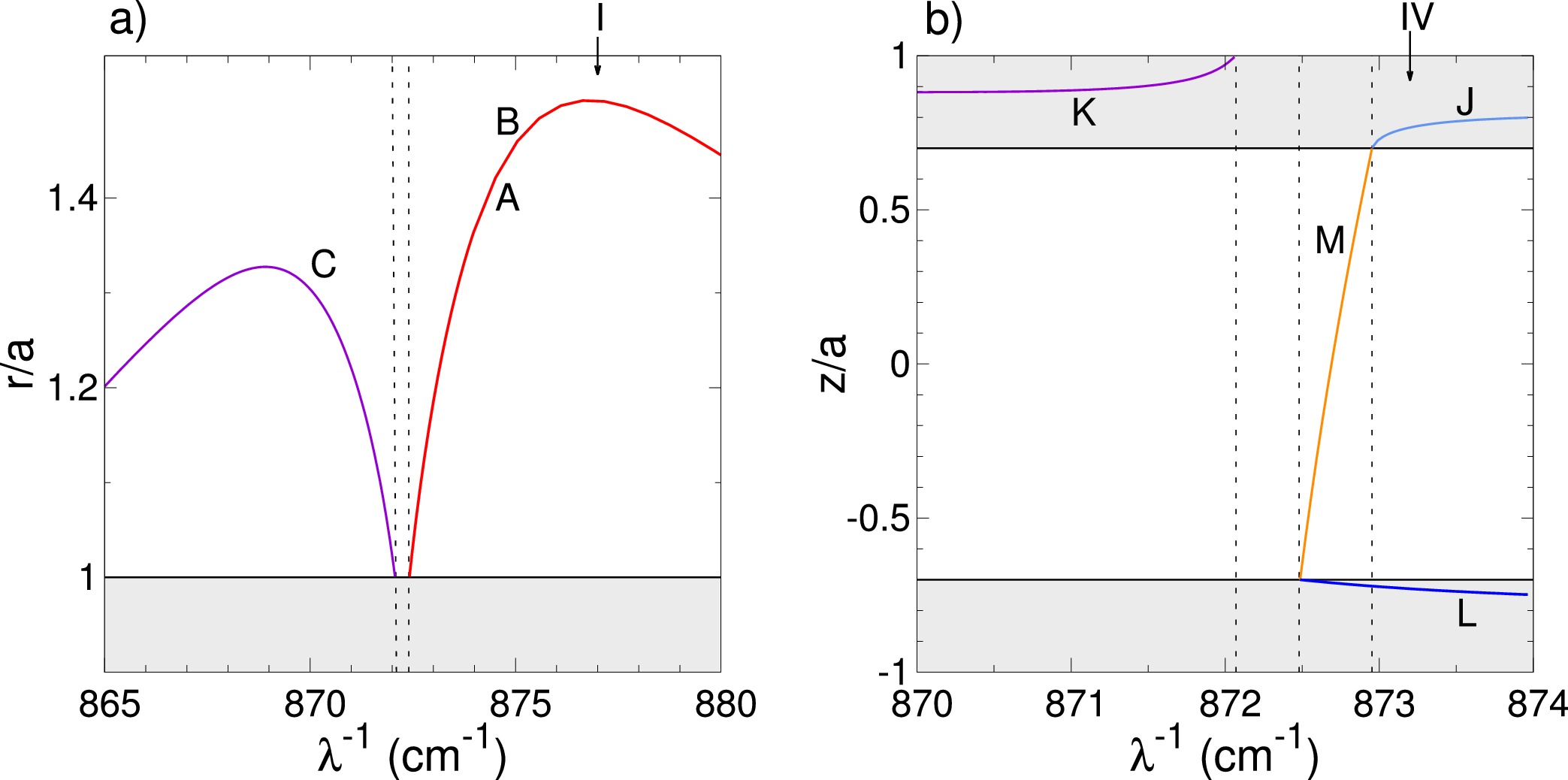}%
 \caption{\label{SRInside_r} (Color online) Position of singular points (Arabic letters) in the 
 Poynting field outside (a) and inside (b) a \CaO/\AlTwoOThree\ particle with $f=0.7$ and  
 $a=0.4\,\mu{\rm m}$ for wave numbers near the antibonding resonance at 
 $\lambda_{\rm ab,1}^{-1}\approx 871.62\, {\rm cm}^{-1}$. 
 Roman numbers indicate regions with fixed topology which were also found for the model particle.
 The shaded regions denote the shell.}
\end{figure*}

We now go through Fig.~\ref{SRInside} step by step. The antibonding dipole resonance is at 
$\lambda_{\rm ab,1}^{-1}\approx 1767.293\,{\rm cm}^{-1}$. A representative topology of the Poynting
vector field for $\lambda^{-1}<\lambda_{\rm ab,1}^{-1}$ is the one of region VIII shown in 
the upper right panel of Fig.~\ref{SRInside} for the particular wave number 
$\lambda^{-1}= 1767.289\,{\rm cm}^{-1}$. Like before the singular points are the intersection points 
between two lines of different color corresponding respectively to the $S_r=0$ and $S_{\theta}=0$ 
isocline. Very close to horizontal $S_r=0$ isoclines (blue) in the negative $xz$ half-plane is also 
a $S_{\theta}=0$ isocline (green). Both come from the outside and have presently no physical consequences. 
In region VIII there are two singular points inside the shell, K and J, and, as discussed before
(see Fig.~\ref{SRaussen}), two singular points outside the particle, C and D, from which only D is within 
the range of Fig.~\ref{SRInside}. 
The Poynting flux shows that the particle is emitting and reabsorbing light to-and-fro the outside region. 
In the vicinity of the core the Poynting flux in the shell resembles the flux around an excited 
dipole. The core is emitting energy into the shell and is also reabsorbing energy from it. The 
singular points inside the shell arise therefore from an interference between the electromagnetic 
fields of an inner and an outer dipole.

In region VIII the saddle point K arises from field lines emitted from and reabsorbed by the 
outer dipole meeting field lines which are emitted from the inner dipole. The field lines 
take then their way to the left and right, that is, to $\pm x$, causing thereby a rectified 
energy flux. At the saddle point J, on the other hand, the rectified energy flux is divided, 
some field lines are reentering the core and some are leaving the particle. In the shell region,
the points J and K operate thus analogously to the points C and D in the outside region of the 
particle (cp. region VIII of Fig.~\ref{SRaussen}). One can hence interpret the reabsorbed 
energy flow of the outer dipole as light which is incident on the core and interacting there
with the inner dipole. 

For wave numbers closer to the resonance the outside saddle point D enters the shell region, 
as can be seen in the top left panel of Fig.~\ref{SRInside}, where for 
$\lambda^{-1}\approx 1767.29\,{\rm cm}^{-1}$ the green line corresponding to the singular point 
D is coming up form $z/a=-1$. In the top left panel of Fig.~\ref{SRaussen}, on the other hand, 
the green line disappears for that wave number at $r/a=1$. 
The point D is caused by the intersection of the $S_r=0$ isocline (blue line)
with the $S_ \theta=0$ isocline (green line), which lies on the  $z$-axis. With the point
D the $S_r=0$ isocline is also entering the particle. There has to be thus an avoided
crossing with the $S_r=0$ isocline already existing inside the shell because isoclines are 
solutions of the differential equation $\frac{\mathrm{d}r}{\mathrm{d}\theta}=r\frac{S_r}{S_\theta}$ 
and a crossing would violate the uniqueness of the solution.~\cite{WLH04} As a result,
see the top left panel of Fig.~\ref{SRInside}, there is only one singular point--the point 
K--in region VIIb, where  $\lambda^{-1}<\lambda_{\rm ab,1}^{-1}$, and one singular point--the 
point L--in region Vb, where $\lambda^{-1}>\lambda_{\rm ab,1}^{-1}$. The Poynting flux for the
latter is shown in the lower right panel of Fig.~\ref{SRInside} and will be discussed in a 
moment.

\begin{figure*}
 \includegraphics[width=\linewidth]{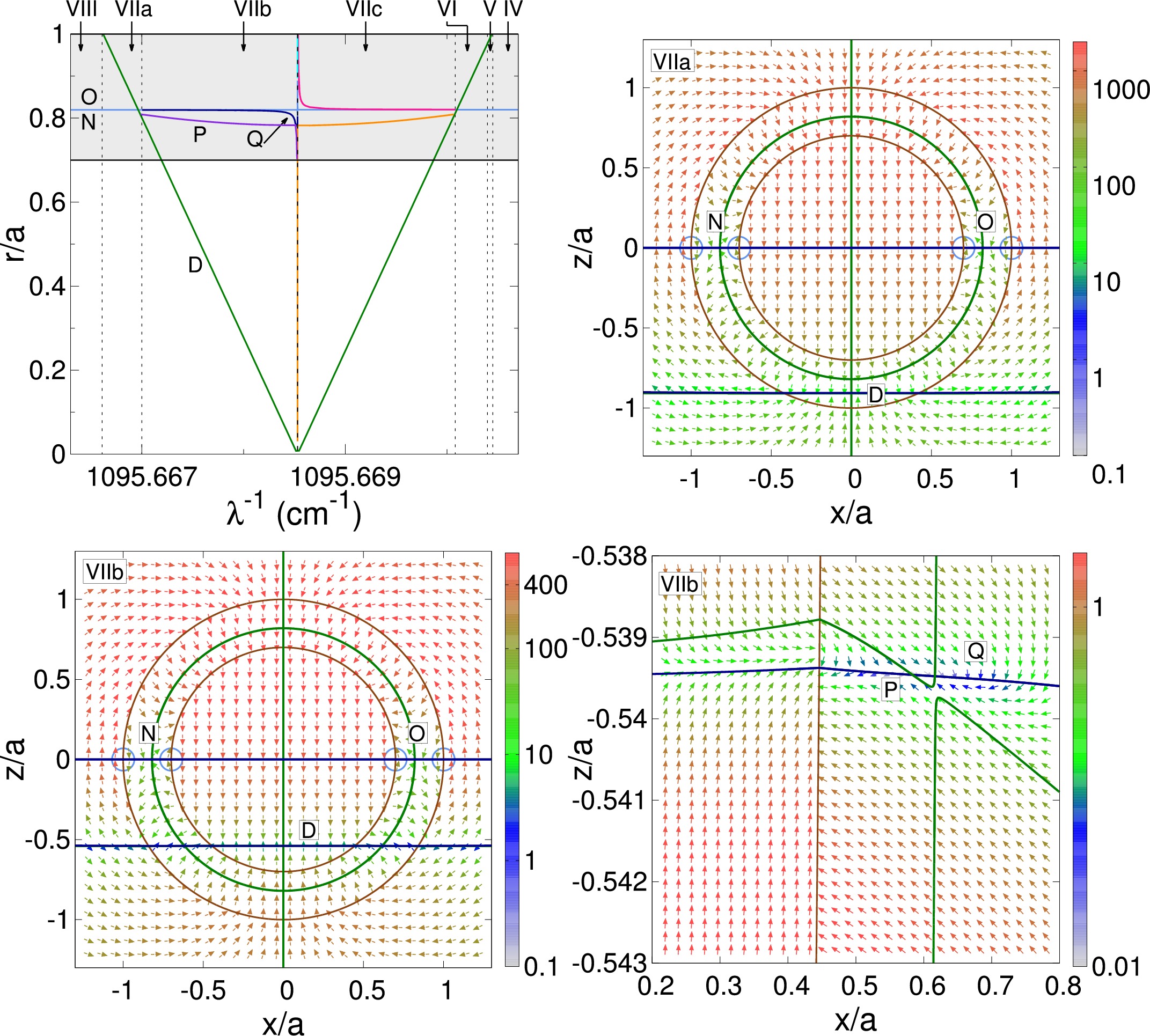}%
 \caption{\label{ARInside} (Color online) Top left: Position of the inside singular points (Arabic 
 letters) in the Poynting field of the model particle ($f=0.7$ and $a=0.1\,\mu{\rm m}$) in the vicinity 
 of the bonding resonance at $\lambda_{\rm b,1}^{-1}\approx 1095.669\,{\rm cm}^{-1}$. Roman 
 numbers indicate spectral regions with a fixed topology while the shaded area denotes the shell. 
 The upper right and the lower left panel show the Poynting field for two representative wave 
 numbers whereas the panel on the lower right is just a zoom-in into the panel to its left. 
 The modulus of the Poynting vector normalized to the flux of incident energy is
 encoded in the color of the arrows. The intersections of the isoclines $S_\theta(r, \theta)=0$ 
 (green) and $S_r(r, \theta)=0$ (blue) determine the singular points. The shell and the core
 are indicated by brown circles and the optical whirlpools are shown as light blue circles.}
\end{figure*}

The avoided crossing occurs also in region VIIc the Poynting flux of it is plotted in the 
bottom left panel of Fig.~\ref{SRInside}. The $S_r=0$ isocline entering the shell from the 
outside goes around it taking its way through the upper half of the $xz$-plane and causing 
thereby the saddle point K. While the $S_r=0$ isocline inside the shell 
enters the core to prevent a crossing and causes thereby the saddle-node point M inside the 
core and the saddle point L inside the shell. The saddle point L has the same function as 
the saddle point K. The emitted field lines of the outer dipole are reabsorbed in the shell
and meet there the emitted field lines of the inner dipole. The saddle point M, on the other
hand, distinguishes between a region where both the inner and the outer dipole have 
already changed their directions from a region where they have not. 
In the $xz$-plane this point is a singularity line. Thus to visualize the 
Poynting flux around this saddle point one would have to look at the $yz$-plane.

The topology of the Poynting flux in region Vb is plotted in the lower right panel of 
Fig.~\ref{SRInside} for $\lambda^{-1}=1767.2965\,$cm$^{-1}$. The singular point L can 
be clearly seen. Since $\lambda^{-1}>\lambda^{-1}_{\rm ab,1}$ both dipoles have changed their 
direction as can be also inferred from the orientation of the energy whirlpools at $z=0$. 
For wave numbers well above $\lambda^{-1}_{\rm ab,1}$ the topology of the energy flux inside 
the particle is the one of region IV. There are now two saddle points L and J corresponding 
to the singular points K and J in region VIII, respectively. The topology before and after 
the resonance is thus in fact the same only upside down because the inner and the outer 
dipole changed simultaneously their direction.

Up to this point we discussed the Poynting flux near the antibonding dipole resonance of the 
dissipationless model particle. Since the dielectric functions for the core and the shell 
are real (see, Eq.~\eqref{EpsModel}) only radiative damping is thus accounted for so far. 
In a real particle, however, dissipative losses occur as well. The dielectric functions 
for the core and the shell of the \CaO/\AlTwoOThree\ particle plotted in the left bottom 
panel of Fig.~\ref{shellres} have imaginary parts. From the work of Wang and coworkers~\cite{WLH04} 
we know that the topology of the Poynting vector is very sensitive to dissipation. We expect
therefore the Poynting flux of the \CaO/\AlTwoOThree\ particle to deviate from the Poynting flux
of the model particle. Indeed for the antibonding resonance we found by artificially adding 
an imaginary part to the dielectric function of the model particle that large imaginary parts 
of the core and the shell dielectric functions lead in the shell region to a dominance of the 
inner dipole and to a strong damping of the outer dipole. Since the saddle points are an 
outgrowth of the relative strength of the two dipoles it is clear that not all the singular 
points of the dissipationless model particle may appear for a particle with dissipation. For 
the particular case of a \CaO/\AlTwoOThree\ particle this can be seen in Fig.~\ref{SRInside_r}, 
where we plot for wave numbers near the antibonding dipole resonance in panel (a) the positions 
of the outer singular points and in panel (b) the positions of the inner singular points. Compared 
to the model particle many outer singular points are missing, while most of the inner singular 
points are still present.

\subsubsection{Bonding dipole resonance}

\begin{figure*}
 \includegraphics[width=0.9\linewidth]{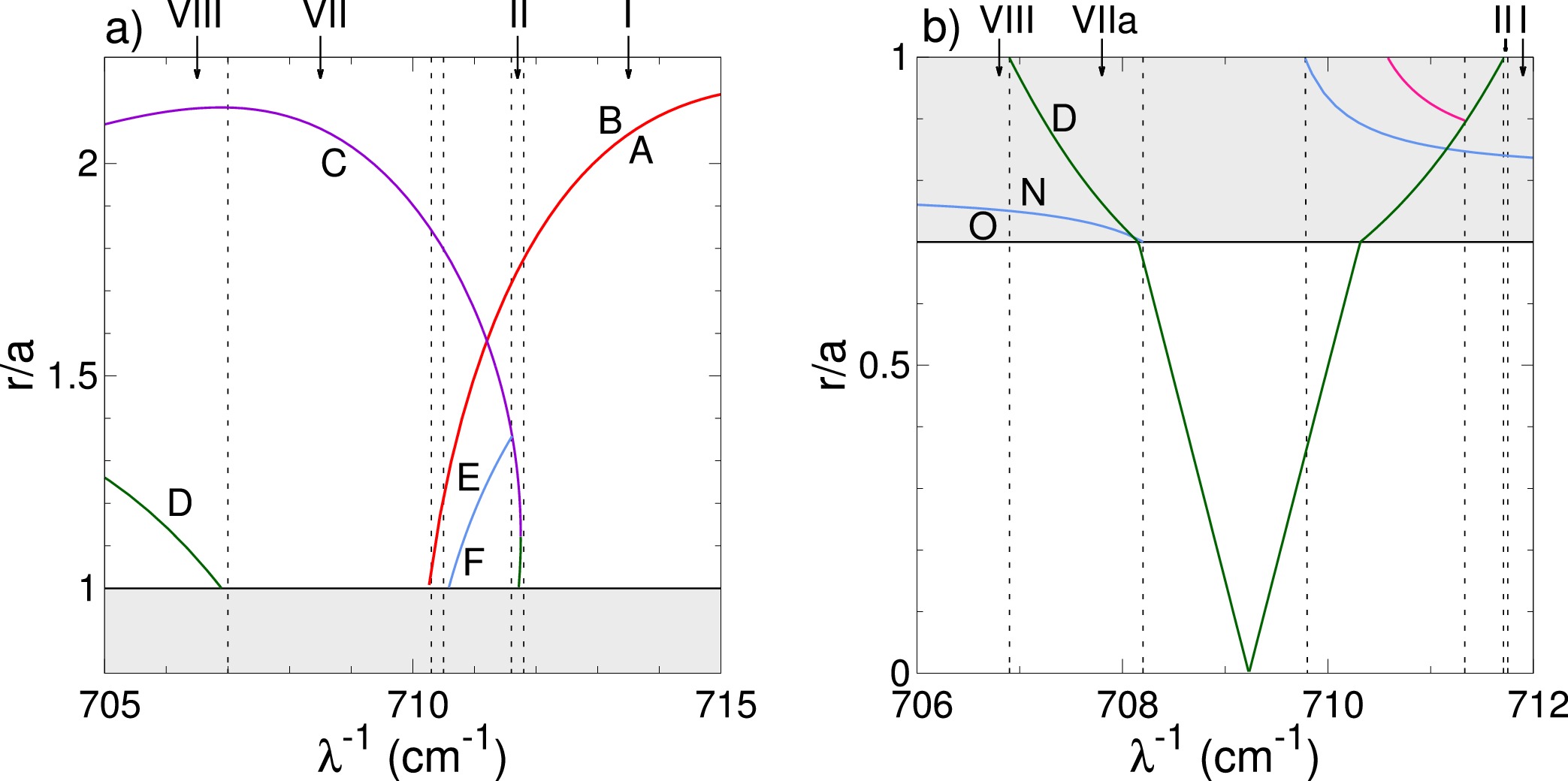}%
 \caption{\label{ARInside_r} (Color online) Position of the singular points (Arabic letters) 
 of the Poynting field outside (a) and inside (b) of a \CaO/\AlTwoOThree\ particle with 
 filling factor $f=0.7$ and radius $a=0.4\,\mu{\rm m}$ in the vicinity of the bonding 
 resonance at $\lambda^{-1}_{\rm b,1}\approx 709.63\,{\rm cm}^{-1}$. Roman numbers indicate  
 regions with a fixed topology which were also found for the model particle and the 
 shaded regions denote the shell.}
\end{figure*}

Let us return to the model particle and discuss the topology of the Poynting vector field near the
bonding dipole resonance at $\lambda_{\rm b,1}^{-1}\approx 1095.669\, {\rm cm}^{-1}$. The results 
are summarized in Fig.~\ref{ARInside}. As it was the case for the antibonding resonance, optical whirlpools 
appear at the core-shell and the shell-vacuum interface. They correspond to surface localized electromagnetic 
modes. However, since at the bonding resonance the inner and the outer dipole are oriented parallel, the 
whirlpools at the two interfaces rotate now in the same direction. For $\lambda^{-1}<\lambda^{-1}_{\rm b,1}$, 
for instance, in region VIIa, the whirlpools at $x/a=-1, z/a=0$ and $x/a=-0.7, z/a=0$ both rotate clockwise 
while the whirlpools at $x/a=0.7, z/a=0$ and $x/a=1, z/a=0$ both rotate anti-clockwise. Behind the resonance, 
that is, for $\lambda^{-1}>\lambda^{-1}_{\rm b,1}$, for instance, in region VIIc the situation is reversed.

In the vicinity of two rectified optical whirlpools energy flows in the opposite direction. As a result 
the singular points N and O appear in region VIIa and region VIIb. They are caused by a $S_{\theta}=0$ 
isocline inside the shell, indicated by a green line, which forms a circle around the core. The two singular 
points N and O, which separate the field lines entering the core from the field lines leaving the particle, 
operate thus in the same way as the two singular points A and B of region I outside the particle (see 
lower right panel in Fig.~\ref{SRaussen}) which separate the field lines entering the particle from the
ones leaving it.

In addition to the points N and O for wave numbers close to the resonance the saddle point D appears
inside the particle. The closer the wave number is to the resonance the more moves D inside the particle. 
In region VIIa D is inside the shell while in region VIIb D is inside the core. Underneath the $S_{r}=0$ 
isocline at $z<0$ is moreover hidden another $S_{\theta}=0$ isocline which is coming from the outside. 
Hence, if the saddle point D is deep enough inside the particle an avoided crossing occurs between the 
two $S_{\theta}=0$ isoclines. In the lower left panel of Fig.~\ref{ARInside} only the $S_{\theta}=0$
isocline encircling the core is visible. To clarify the situation we replotted it therefore on a 
magnified scale in the lower right panel. There it can be seen that the 
$S_{\theta}=0$ isocline coming from the outside of the particle is going downwards in the shell,
while the other $S_{\theta}=0$ isocline is going towards the core. As a result the singular points 
Q and P appear. While the point Q is an unstable foci, that is, an energy sink in the $xz$-plane, the 
point P is a saddle point. In region VI, where $\lambda^{-1}>\lambda^{-1}_{b,1}$, the saddle point D 
is still inside the shell but lies now above the $S_{\theta}=0$ isocline existing in the shell and 
encircling the core. The topology in this region is the same as in region VIIa, but upside 
down and with different orientation of the optical whirlpools, since the inner and outer dipole 
simultaneously changed directions.

Having discussed for the dissipationless model particle the topology of the inside Poynting vector 
field near the bonding dipole resonance, we now comment on the energy flux near such a resonance in 
a real particle with finite dissipation, taking again the \CaO/\AlTwoOThree\ particle as an example. 
For the bonding resonance we empirically found by varying the imaginary parts of the dielectric 
function the outer dipole to be rather stable, even for large values of $\epsilon''_{1,2}$. 
The inner dipole however disappears. As a result there are also no surface localized electromagnetic 
modes at the core-shell interface anymore. The results for the bonding dipole resonance of the 
\CaO/\AlTwoOThree\ particle shown in Fig.~\ref{ARInside_r} are consistent with these empirical findings. 
Whereas almost all singular points in the Poynting flux outside the particle are preserved because of 
the robustness of the outer dipole, inside the shell most of the singularities disappeared. The 
situation is thus just reversed to the one we found for the antibonding resonance.

\subsection{Electric field} \label{Efield}

We now investigate for the model particle the electric field for wavelengths in the vicinity of the 
bonding and antibonding dipole resonance and contrast it with the field in the vicinity of the core 
resonance. For comparison we also include the field near the anomalous resonance of an homogeneous
particle. For all four resonances we calculate the electric field amplitude $|\vec{E}|^2$ and the real 
part of the electric field $\mathrm{Re}(\vec{E})$ for wave numbers where the topology of the Poynting 
flux is of type I shown in Fig.~\ref{SRaussen}: For the homogeneous particle 
$\lambda^{-1}=1406.5\,$cm$^{-1}$, while for the core-shell particle $\lambda^{-1}=306.026\,$cm$^{-1}$ (core 
resonance), $\lambda^{-1}=1095.74\,$cm$^{-1}$ (bonding resonance), and $\lambda^{-1}=1767.375\,$cm$^{-1}$ 
(antibonding resonance). 

\begin{figure*}
 \includegraphics[width=\linewidth]{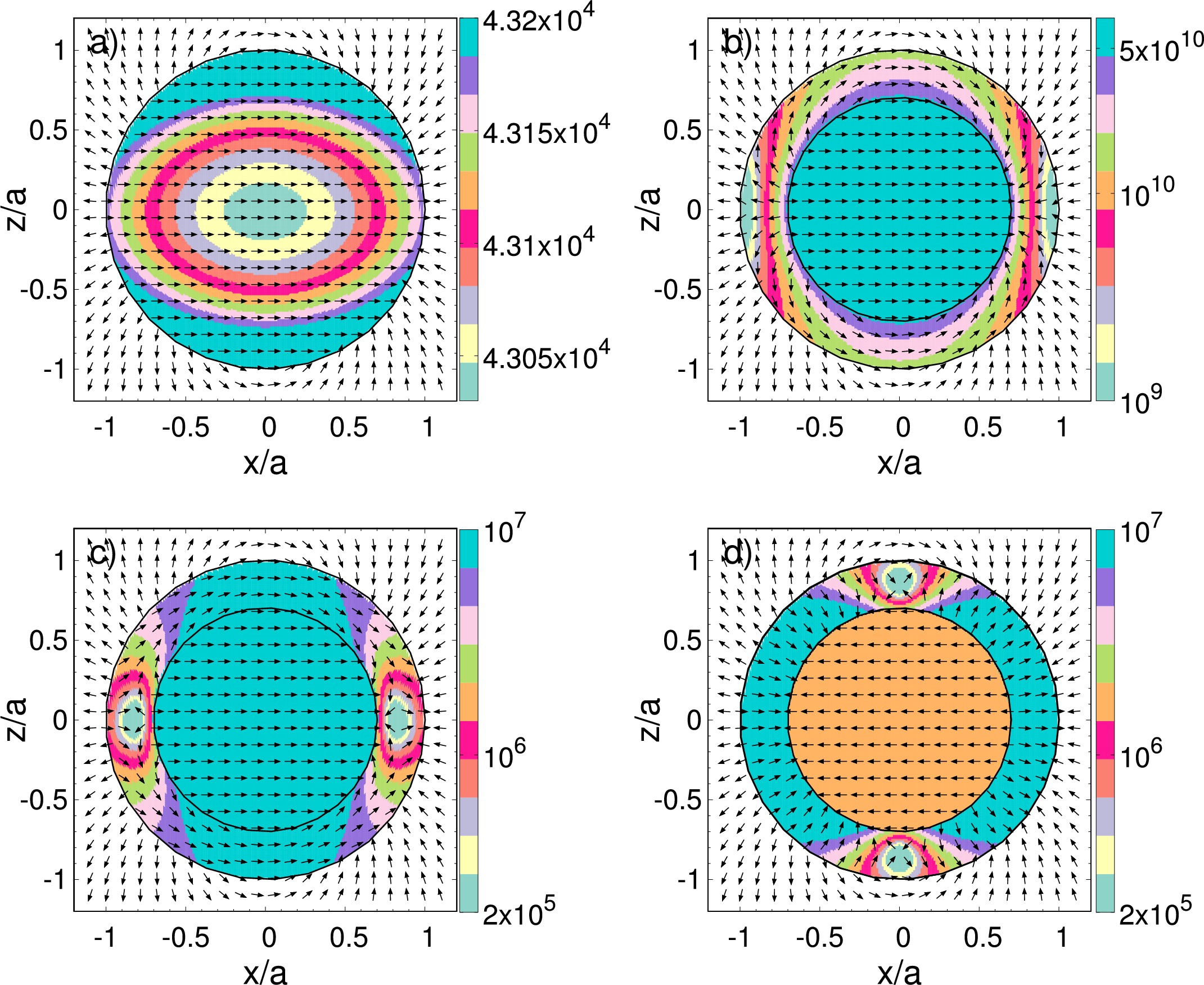}%
 \caption{\label{Eqaud} (Color online) Electric field amplitude $|\vec{E}|^2$ and the direction of the real 
  part of the electric field in the vicinity of the core (b), the bonding (c), and the antibonding (d) dipole
  resonance of the model particle ($f=0.7$ and $a=0.1\,\mu{\rm m}$). Panel (a) shows for comparison
  results for the dipole resonance of a homogeneous particle of the same size made out of the core
  material. Black arrows denote the direction of the real part of the electric field while the colors
  give the field amplitude. The wave numbers used for the plots are given in the main text.}
\end{figure*}

The results for the electric field are summarized in Fig.~\ref{Eqaud}, where the 
value of the electric field amplitude is encoded in colors and the direction of the real part of the electric 
field is given by black arrows. The electric field of the incident wave is polarized in $x$-direction. 
Hence, the induced charge displacements are also in $x$-direction. 

Let us first discuss the field for the anomalous resonance of the homogeneous particle shown in 
panel (a) of Fig.~\ref{Eqaud}. From the electric field structure outside the particle it is clear 
that a dipole mode is excited. The field lines by definition go from positive to negative charges 
but the dipole of course points towards positive charges. Hence, the excited dipole is oriented in 
negative $x$-direction implying polarization-induced surface charges to be positive for the $x<0$ 
hemisphere and negative for the $x>0$ hemisphere. As can be seen from the linear scale of the 
plot, the variation of the electric field amplitude inside the homogeneous particle is very small.
Yet, a shallow minimum occurs at the center of the particle. 

The electric field for the core-shell particle is plotted in panels (b) to (d) 
for the core, the bonding, and the antibonding resonance, respectively. Inside the core the electric 
field resembles for all three resonances the field of an homogeneous particle. On the logarithmic 
scale used for the plots this cannot be seen but a linear scale would reveal an onion-like structure
similar to the one depicted in panel (a). Only for the core resonance (b) however would the minimum of 
the field still be at the center of the particle whereas for the bonding and antibonding resonance 
the field at the center is in fact maximal. The core resonance (b) has most in common with the 
anomalous resonance of the homogeneous particle (a) because it arises also from the resonance 
condition of the core dielectric function. Only a single dipole is thus excited as can be seen from 
the field lines. They originate from the $x<0$ hemisphere of the core-shell interface and terminate 
in its $x>0$ hemisphere, signalling positive and negative polarization-induced surface charges, 
respectively. At the shell-vacuum interface the field lines are only refracted leading to an 
inhomogeneous field inside the shell which is in most parts much smaller than in the core.

For the bonding and antibonding resonance the situation is different because now two dipoles are 
involved. From the polarization-induced surface charges shown in Fig.~\ref{SurfaceMode} we already 
know that for the bonding resonance the dipoles are orientated parallel whereas for the antibonding 
resonance they are oriented antiparallel. The field lines shown in the panels (c) and (d) verify this. 
Let us first look at panel (c) displaying the field for the bonding dipole resonance. The shell-core 
interface in the $x<0$ hemisphere is a source for field lines, indicating positive charges, whereas
the shell-core interface in the $x>0$ hemisphere is a sink for field lines, indicating negative 
charges. At the shell-vacuum interface the situation is the same. Both the inner and the outer dipole
are thus oriented in negative $x$-direction. Inside the shell the two dipole fields interfere which
destroys the homogeneity of the field. For $z=0$ and $x/a \approx \pm 0.9$ destructive interference 
occurs as the outgoing field lines from the inner dipole cancel the inner field lines of the outer dipole. 
At $z/a=\pm 0.9$ and $x=0$, on the other hand, constructive interference occurs, because outer field 
lines of the inner dipole and inner field lines of the outer dipole run parallel leading to a high 
electric field. At the antibonding resonance shown in panel (d) the shell-core interface in the $x<0$ 
hemisphere is a sink for field lines while in the $x>0$ hemisphere it is a source. The shell-vacuum 
interface on the other hand is a source for field lines in the $x<0$ hemisphere and a sink $x>0$ 
hemisphere. The inner and the outer dipole are thus oriented antiparallel. In the shell the inner 
field lines of the outer dipole interfere with the outer field lines of the inner dipole, leading at 
$z/a=\pm 0.9$ and $x/a=0$ to a minimum in the field.

The electric field in core-shell particles differs therefore dramatically from the one in an 
homogeneous particle. Whereas the latter is homogeneous the former is very inhomogeneous, varying 
in the shell for the bonding and antibonding resonance over almost two orders of magnitude. At the 
bonding resonance the field amplitude is also high in the core but for the antibonding resonance it 
is one order of magnitude smaller in the core than in most parts of the shell. The inhomogeneity of the 
electric field can be used to control the dissipation of energy inside a core-shell particle by varying 
the wavelength of the incident light. For the model particle discussed in this subsection there is of 
course no dissipation since the dielectric functions of the core and the shell are purely real but for 
physical particles the dielectric functions are complex and a control mechanism is conceivable. 

\subsection{Dissipation}

In this section we present results for the dissipation of energy inside a \CaO/\AlTwoOThree\ particle.
According to Eq.~\eqref{eq:dissip} it depends on the imaginary parts of the dielectric functions 
as well as on the amplitude of the electric field. Since the latter is strongly enhanced in the vicinity 
of anomalous resonances, even small imaginary parts of the dielectric functions may thus lead to a huge 
dissipation.~\cite{LMT12} The field enhancement in a core-shell particle is moreover rather inhomogeneous, 
there will be hence spots inside it where dissipation is strong and spots where it will be rather weak, 
irrespective of the magnitude of the (spatially homogeneous) imaginary parts of the dielectric functions.

\begin{figure*}
 \includegraphics[width=\linewidth]{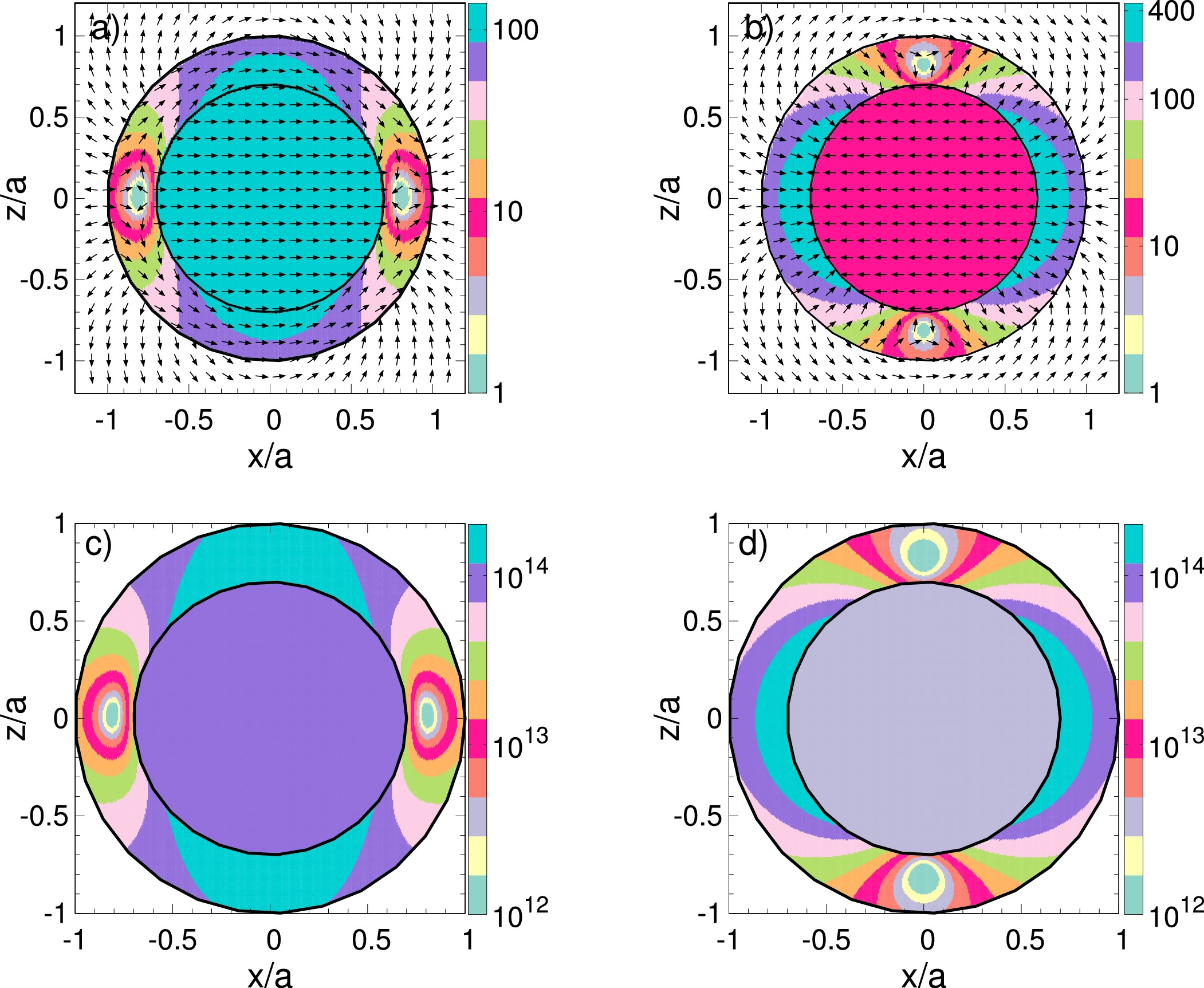}%
 \caption{\label{Eqaud_dissip_real} (Color online) The top two panels show the electric field amplitude 
  $|\vec{E}|^2$ and the direction of the real part of the electric field in the vicinity of the bonding 
  (a) and the antibonding (b) dipole resonance of a \CaO/\AlTwoOThree\ particle with $f=0.7$ and 
  $a=0.4\,\mu {\rm m}$. The wave numbers are respectively $\lambda^{-1}=715\,{\rm cm}^{-1}$ 
  and $\lambda^{-1}=877\,{\rm cm}^{-1}$. The two bottom panels depict the dissipation inside the particle 
  for the bonding (c) and antibonding (d) dipole resonance, respectively. Black arrows in the 
  upper two panels denote the direction of the real part of the electric field while colors indicate 
  the electric field amplitude (upper two panels) or the dissipation (lower two panels). }
\end{figure*}

In Fig.~\ref{Eqaud_dissip_real} we show the electric field and the dissipation for the \CaO/\AlTwoOThree\ 
particle near the bonding and the antibonding dipole resonance. The wave numbers are in both cases 
in the topological region I where the Poynting flux outside the particle displays Tribelsky 
ears. For the bonding resonance $\lambda^{-1}=715\,$cm$^{-1}$ leading to $\varepsilon''_{1}=0.104$ and 
$\varepsilon''_{2}=0.304$, while for the antibonding resonance $\lambda^{-1}=877\,$cm$^{-1}$ giving 
rise to $\varepsilon''_{1}=0.0492$ and $\varepsilon''_{2}=0.0657$.

The electric field inside the \CaO/\AlTwoOThree\ particle shown in the upper two panels of 
Fig.~\ref{Eqaud_dissip_real} is inhomogeneous as it was for the model particle discussed in the 
previous section. Due to constructive and destructive interference of the electric fields 
of the excited dipoles inside the shell there are points where the amplitude of the electric field is 
enhanced and points where it is very small. For the bonding resonance (panel (a)) the field amplitude 
inside the core is high and more or less homogeneous whereas inside the shell the field amplitude 
varies over two orders of magnitude reaching at $x/a=\pm 0.9$ and $z/a=0$ its smallest value which 
is two orders of magnitude lower than inside the core. Compared to the dissipationless model particle 
the electric field inside the core of the \CaO/\AlTwoOThree\ particle is no longer symmetric along the 
$x$-axis and it reaches its maximum no longer in the center of the particle but around $z/a=0.7$ and 
$x/a=0$. On the logarithmic scale this cannot be seen but a linear scale would reveal these features.
The spatial distribution of the dissipation near the bonding resonance (panel (c)) parallels more or 
less the field distribution. It is homogeneous inside the core, where it is also rather high, and varies 
strongly inside the shell reaching its smallest value at $x/a=\pm 0.9$ and $z/a=0$. The results for 
the antibonding resonance, given in panels (b) and (d) of Fig.~\ref{Eqaud_dissip_real}, show the 
same overall trend. Inside the core the field and the dissipation are homogeneous whereas in the 
shell strong variations occur. In contrast to the bonding resonance the electric field inside the
core is now however smaller than in most parts of the shell where destructive and constructive 
interference lead moreover to a strongly varying field. This feature is also preserved in the spatial 
distribution of the dissipation. It varies strongly within the shell and is in most parts of it 
two orders of magnitude higher than inside the core. There are however also narrow regions where 
dissipation is strongly suppressed.  

The spatial inhomogeneities of the electric field in the shell of the model particle can thus be also 
found in the shell of the \CaO/\AlTwoOThree\ particle. Since the dielectric functions of the latter 
have finite imaginary parts, the field inhomogeneities lead now however also to inhomogeneities in 
the dissipation. Within the shell the dissipation of energy varies for both the bonding and the
antibonding resonance over almost two orders of magnitude giving rise to extended regions of high 
dissipation and well localized cold spots where dissipation is rather low although the dielectric
function of the shell is spatially homogeneous. 

\section{Conclusions}
\label{Conclusions}

In our previous work we proposed to use the antibonding dipole resonance of dielectric core-shell 
particles having a core with negative and a shell with positive electron affinity to determine
by infrared attenuation spectroscopy the number of surplus electrons of the particle. 
Due to the particular choice of the electron affinities the electrons would be confined to the 
shell region leading to a high volume electron density which in turn would strongly affect the electric 
polarizability of the particle and hence the position of the extinction resonances. In particular 
the maximum of the antibonding dipole resonance, which we initially called shell resonance since 
it is absent for homogeneous particles, turned out to be very charge sensitive. 
To understand the mechanism leading to this resonance and also to make contact to the field
of plasmonics, we investigated in the present work neutral dielectric core-shell particles, using a 
\CaO/\AlTwoOThree\ particle as a physical example and a dissipationless model core-shell particle 
as an idealized system. The scattering of light is in both cases mediated by infrared transverse optical 
phonons. 

We showed that the series of bonding and antibonding resonances arising in the surface mode 
regime of the shell are attached to in-phase and out-of-phase multipolar polarization-induced surface 
charges building up at the two interfaces to satisfy the boundary conditions of the Maxwell equations. 
For the series of bonding resonances the surface charges are in-phase whereas for the series of 
antibonding resonances they are out-of-phase. The inner and outer dipoles associated with the two 
lowest resonances are thus aligned parallel for the bonding and antiparallel for the antibonding 
resonance. We demonstrated that both series of resonances scatter light anomalously, that is, originate
from selective excitation of surface modes giving then rise to an inverse hierarchy in the extinction 
cross section. We also analyzed the topology of the Poynting flux in the vicinity of the bonding 
and the antibonding dipole resonance. The outside Poynting flux near these two resonances is determined 
by an effective dipole describing the effective outer field of the inner and outer dipole. 
The topology of the outside Poynting flux is thus for the bonding and antibonding resonance 
similar to the Poynting flux of an homogeneous particle. Inside the particle however 
the Poynting flux is determined by the interplay of the outer field of the inner and the inner 
field of the outer dipole. As a result, the topology of the inside Poynting flux depends on the 
alignment of the two dipoles and is thus different for the bonding and antibonding resonance.  

Inside the particle the Poynting flux, the electric field, and the dissipation of energy are highly
inhomogeneous. Of particular interest, perhaps also from the technological view point of laser 
heating, is the inhomogeneous dissipation near the antibonding dipole resonance. Inside the shell 
there are broad banana-shaped hot spots where the dissipation is almost two orders of magnitude 
larger than in the core as well as rather localized cold spots where the dissipation is almost 
an order smaller. Although we do not expect the enhanced dissipation in the shell to be a problem 
for the optical charge measurement we proposed since the surplus electrons are bound by an 
energy on the order of the shell material's electron affinity, which is for technologically relevant 
dielectrics such as \AlTwoOThree\ in the range of electron volts, the heating of the confined 
electron gas should be investigated in the future.

\begin{acknowledgments}
This work was supported by the Deutsche Forschungsgemeinschaft through the Transregional 
Collaborative Research Center SFB/TRR24. 
\end{acknowledgments}

\section*{Appendix}

Using the orthogonality of the vector spherical harmonics,~\cite{BH83,Stratton41} the eight
expansion coefficients $a_n$, $b_n$, $c_n$, $d_n$, $f_n$, $g_n$, $v_n$ and $w_n$ can be
straightforwardly calculated from the core-shell boundary conditions at $r=b$, 
\begin{align}
\big(\vec{E}_2-\vec{E}_1\big)\times\vec{r} &=0~,\\
\big(\vec{H}_2-\vec{H}_1\big)\times\vec{r} &=0~,
\end{align}
and the shell-vacuum boundary conditions at $r=a$
\begin{align}
\big(\vec{E}_s+\vec{E}_{\rm in}-\vec{E}_2\big)\times\vec{r} &=0\label{BChom1}~,\\
\big(\vec{H}_s+\vec{H}_{\rm in}-\vec{H}_2\big)\times\vec{r} &=0~,
\label{BChom2}
\end{align}
yielding eight equations for the eight unknowns. 

The coefficients $a_n$ and $b_n$ determining the scattering components of the electromagnetic fields 
in the outer space, Eqs.~\eqref{Es}--\eqref{Hs}, turn then out to be~\cite{AK51,BH83}  
\begin{align}
a_n &= C_n^{-1}\bigg\{\psi_n(y)[\psi_n^\prime(m_2y)-A_n\chi_n^\prime(m_2y)]\nonumber\\
&- m_2\psi_n^\prime(y)[\psi_n(m_2y)-A_n\chi_n(m_2y)]\bigg\}~,\label{acoeff}\\
b_n &= D_n^{-1}\bigg\{m_2\psi_n(y)[\psi_n^\prime(m_2y)-B_n\chi^\prime(m_2y)]\nonumber\\
&- \psi_n^\prime(y)[\psi_n(m_2y)-B_n\chi_n(m_2y)]\bigg\}\label{bcoeff}
\end{align}
with
\begin{align}
A_n &= \frac{m_2\psi_n(m_2x)\psi_n^\prime(m_1x)-m_1\psi_n^\prime(m_2x)\psi_n(m_1x)}
{m_2\chi_n(m_2x)\psi_n^\prime(m_1x)-m_1\chi_n^\prime(m_2x)\psi_n(m_1x)}~,\\
B_n &= \frac{m_2\psi_n(m_1x)\psi_n^\prime(m_2x)-m_1\psi_n(m_2x)\psi_n^\prime(m_1x)}
{m_2\chi_n^\prime(m_2x)\psi_n(m_1x)-m_1\psi_n^\prime(m_1x)\chi_n(m_2x)}~,\\
C_n &= \xi_n(y)[\psi_n^\prime(m_2y)-A_n\chi_n^\prime(m_2y)]
-m_2\xi_n^\prime(y)\nonumber\\
&\times[\psi_n(m_2y)
-A_n\chi_n(m_2y)]~,\\
D_n &= m_2\xi_n(y)[\psi_n^\prime(m_2y)-B_n\chi_n^\prime(m_2y)]
- \xi_n^\prime(y)\nonumber\\
&\times[\psi_n(m_2y)-B_n\chi_n(m_2y)]~.
\end{align}
For the fields inside the core, Eqs.~\eqref{Ecore}--\eqref{Hcore}, the coefficients read 
\begin{align}
c_n &= -D_n^{-1} F_n m_2 m_1\nonumber\\
& \times \frac{[\psi_n(m_2x)\chi_n^\prime(m_2x) - \chi_n(m_2x)\psi_n^\prime(m_2x)]}{[m_1\chi_n(m_2x)\psi_n^\prime(m_1x) 
- m_2\psi_n(m_1x)\chi_n^\prime(m_2x)]}~,\label{ccoeff}\\
d_n &=-C_n^{-1} F_n m_2 m_1\nonumber\\
& \times \frac{[\psi_n(m_2x)\chi_n^\prime(m_2x) - \chi_n(m_2x)\psi_n^\prime(m_2x)]}{[m_2\chi_n(m_2x)\psi_n^\prime(m_1x) 
- m_1\psi_n(m_1x)\chi_n^\prime(m_2x)]}\label{dcoeff}
\end{align}
with
\begin{align}
F_n = [\psi_n^\prime(y)\xi_n(y) - \psi_n(y)\xi_n^\prime(y)]~,
\end{align}
while for the fields inside the shell, Eqs.~\eqref{Eshell}--\eqref{Hshell}, the coefficients become
\begin{align}
f_n &= D_n^{-1} m_2 F_n~,\label{fcoeff}\\
g_n &= C_n^{-1} m_2 F_n~,\label{gcoeff}\\
v_n &= D_n^{-1} m_2 F_n B_n~,\label{vcoeff}\\
w_n &= C_n^{-1} m_2 F_n A_n~.\label{wcoeff}
\end{align}


\begin{thebibliography}{37}
\expandafter\ifx\csname natexlab\endcsname\relax\def\natexlab#1{#1}\fi
\expandafter\ifx\csname bibnamefont\endcsname\relax
  \def\bibnamefont#1{#1}\fi
\expandafter\ifx\csname bibfnamefont\endcsname\relax
  \def\bibfnamefont#1{#1}\fi
\expandafter\ifx\csname citenamefont\endcsname\relax
  \def\citenamefont#1{#1}\fi
\expandafter\ifx\csname url\endcsname\relax
  \def\url#1{\texttt{#1}}\fi
\expandafter\ifx\csname urlprefix\endcsname\relax\def\urlprefix{URL }\fi
\providecommand{\bibinfo}[2]{#2}
\providecommand{\eprint}[2][]{\url{#2}}

\bibitem[{\citenamefont{Oldenburg et~al.}(1998)\citenamefont{Oldenburg,
  Averitt, Westcott, and Halas}}]{OA98}
\bibinfo{author}{\bibfnamefont{S.~J.} \bibnamefont{Oldenburg}},
  \bibinfo{author}{\bibfnamefont{R.~D.} \bibnamefont{Averitt}},
  \bibinfo{author}{\bibfnamefont{S.~L.} \bibnamefont{Westcott}},
  \bibnamefont{and} \bibinfo{author}{\bibfnamefont{N.~J.} \bibnamefont{Halas}},
  \bibinfo{journal}{{Chem. Phys. Lett.}} \textbf{\bibinfo{volume}{288}},
  \bibinfo{pages}{243} (\bibinfo{year}{1998}).

\bibitem[{\citenamefont{Halas et~al.}(2012)\citenamefont{Halas, Lal, Link,
  Chang, Natelson, Hafner, and Nordlander}}]{HLL12}
\bibinfo{author}{\bibfnamefont{N.~J.} \bibnamefont{Halas}},
  \bibinfo{author}{\bibfnamefont{S.}~\bibnamefont{Lal}},
  \bibinfo{author}{\bibfnamefont{S.}~\bibnamefont{Link}},
  \bibinfo{author}{\bibfnamefont{W.-S.} \bibnamefont{Chang}},
  \bibinfo{author}{\bibfnamefont{D.}~\bibnamefont{Natelson}},
  \bibinfo{author}{\bibfnamefont{J.~H.} \bibnamefont{Hafner}},
  \bibnamefont{and}
  \bibinfo{author}{\bibfnamefont{P.}~\bibnamefont{Nordlander}},
  \bibinfo{journal}{Advanced Materials} \textbf{\bibinfo{volume}{24}},
  \bibinfo{pages}{4842} (\bibinfo{year}{2012}).

\bibitem[{\citenamefont{Fan et~al.}(2014)\citenamefont{Fan, Zheng, and
  Singh}}]{FZ14}
\bibinfo{author}{\bibfnamefont{X.}~\bibnamefont{Fan}},
  \bibinfo{author}{\bibfnamefont{W.}~\bibnamefont{Zheng}}, \bibnamefont{and}
  \bibinfo{author}{\bibfnamefont{W.~J.} \bibnamefont{Singh}},
  \bibinfo{journal}{Light: Sci. Appl.} \textbf{\bibinfo{volume}{3}},
  \bibinfo{pages}{1} (\bibinfo{year}{2014}).

\bibitem[{\citenamefont{Prasad}(2004)}]{P04}
\bibinfo{author}{\bibfnamefont{P.~N.} \bibnamefont{Prasad}},
  \emph{\bibinfo{title}{{Nanophotonics}}} (\bibinfo{publisher}{John Wiley \&
  Sons}, \bibinfo{year}{2004}).

\bibitem[{\citenamefont{Loo et~al.}(2005)\citenamefont{Loo, Lowery, Halas,
  West, and Drezek}}]{LL05}
\bibinfo{author}{\bibfnamefont{C.}~\bibnamefont{Loo}},
  \bibinfo{author}{\bibfnamefont{A.}~\bibnamefont{Lowery}},
  \bibinfo{author}{\bibfnamefont{N.}~\bibnamefont{Halas}},
  \bibinfo{author}{\bibfnamefont{J.}~\bibnamefont{West}}, \bibnamefont{and}
  \bibinfo{author}{\bibfnamefont{R.}~\bibnamefont{Drezek}},
  \bibinfo{journal}{Nano Lett.} \textbf{\bibinfo{volume}{5}},
  \bibinfo{pages}{709} (\bibinfo{year}{2005}).

\bibitem[{\citenamefont{Ozbay}(2006)}]{O06}
\bibinfo{author}{\bibfnamefont{E.}~\bibnamefont{Ozbay}},
  \bibinfo{journal}{{Science}} \textbf{\bibinfo{volume}{311}},
  \bibinfo{pages}{189} (\bibinfo{year}{2006}).

\bibitem[{\citenamefont{Liu et~al.}(2012{\natexlab{a}})\citenamefont{Liu,
  Miroshnichenko, Neshev, and Kivshar}}]{LMN12a}
\bibinfo{author}{\bibfnamefont{W.}~\bibnamefont{Liu}},
  \bibinfo{author}{\bibfnamefont{A.~E.} \bibnamefont{Miroshnichenko}},
  \bibinfo{author}{\bibfnamefont{D.~N.} \bibnamefont{Neshev}},
  \bibnamefont{and} \bibinfo{author}{\bibfnamefont{Y.~S.}
  \bibnamefont{Kivshar}}, \bibinfo{journal}{ACS Nano}
  \textbf{\bibinfo{volume}{6}}, \bibinfo{pages}{5489}
  (\bibinfo{year}{2012}{\natexlab{a}}).

\bibitem[{\citenamefont{Liu et~al.}(2012{\natexlab{b}})\citenamefont{Liu,
  Miroshnichenko, Neshev, and Kivshar}}]{LMN12b}
\bibinfo{author}{\bibfnamefont{W.}~\bibnamefont{Liu}},
  \bibinfo{author}{\bibfnamefont{A.~E.} \bibnamefont{Miroshnichenko}},
  \bibinfo{author}{\bibfnamefont{D.~N.} \bibnamefont{Neshev}},
  \bibnamefont{and} \bibinfo{author}{\bibfnamefont{Y.~S.}
  \bibnamefont{Kivshar}}, \bibinfo{journal}{{Phys. Rev. B}}
  \textbf{\bibinfo{volume}{86}}, \bibinfo{pages}{081407(R)}
  (\bibinfo{year}{2012}{\natexlab{b}}).

\bibitem[{\citenamefont{Zhang et~al.}(2012)\citenamefont{Zhang, Jing, Boisvert,
  He, and Wang}}]{ZJB12}
\bibinfo{author}{\bibfnamefont{L.}~\bibnamefont{Zhang}},
  \bibinfo{author}{\bibfnamefont{H.}~\bibnamefont{Jing}},
  \bibinfo{author}{\bibfnamefont{G.}~\bibnamefont{Boisvert}},
  \bibinfo{author}{\bibfnamefont{J.~Z.} \bibnamefont{He}}, \bibnamefont{and}
  \bibinfo{author}{\bibfnamefont{H.}~\bibnamefont{Wang}}, \bibinfo{journal}{ACS
  Nano} \textbf{\bibinfo{volume}{6}}, \bibinfo{pages}{3514}
  (\bibinfo{year}{2012}).

\bibitem[{\citenamefont{Stockman}(2008)}]{Stockman08}
\bibinfo{author}{\bibfnamefont{M.~I.} \bibnamefont{Stockman}},
  \bibinfo{journal}{Nature Photonics} \textbf{\bibinfo{volume}{2}},
  \bibinfo{pages}{327} (\bibinfo{year}{2008}).

\bibitem[{\citenamefont{Al\'{u} and Engheta}(2005)}]{AE05}
\bibinfo{author}{\bibfnamefont{A.}~\bibnamefont{Al\'{u}}} \bibnamefont{and}
  \bibinfo{author}{\bibfnamefont{N.}~\bibnamefont{Engheta}},
  \bibinfo{journal}{{Phys. Rev. E}} \textbf{\bibinfo{volume}{72}},
  \bibinfo{pages}{016623} (\bibinfo{year}{2005}).

\bibitem[{\citenamefont{Tribelsky and Luk'yanchuk}(2006)}]{TL06}
\bibinfo{author}{\bibfnamefont{M.~I.} \bibnamefont{Tribelsky}}
  \bibnamefont{and} \bibinfo{author}{\bibfnamefont{B.~S.}
  \bibnamefont{Luk'yanchuk}}, \bibinfo{journal}{{Phys. Rev. Lett.}}
  \textbf{\bibinfo{volume}{97}}, \bibinfo{pages}{263902}
  (\bibinfo{year}{2006}).

\bibitem[{\citenamefont{Tribelsky}(2011)}]{Tribelsky11}
\bibinfo{author}{\bibfnamefont{M.~I.} \bibnamefont{Tribelsky}},
  \bibinfo{journal}{{Europhys. Lett.}} \textbf{\bibinfo{volume}{94}},
  \bibinfo{pages}{14004} (\bibinfo{year}{2011}).

\bibitem[{\citenamefont{Thiessen et~al.}(2014)\citenamefont{Thiessen, Heinisch,
  Bronold, and Fehske}}]{THB14}
\bibinfo{author}{\bibfnamefont{E.}~\bibnamefont{Thiessen}},
  \bibinfo{author}{\bibfnamefont{R.~L.} \bibnamefont{Heinisch}},
  \bibinfo{author}{\bibfnamefont{F.~X.} \bibnamefont{Bronold}},
  \bibnamefont{and} \bibinfo{author}{\bibfnamefont{H.}~\bibnamefont{Fehske}},
  \bibinfo{journal}{Eur. Phys. J. D} \textbf{\bibinfo{volume}{68}},
  \bibinfo{pages}{98} (\bibinfo{year}{2014}).

\bibitem[{\citenamefont{Heinisch
  et~al.}(2012{\natexlab{a}})\citenamefont{Heinisch, Bronold, and
  Fehske}}]{HBF12}
\bibinfo{author}{\bibfnamefont{R.~L.} \bibnamefont{Heinisch}},
  \bibinfo{author}{\bibfnamefont{F.~X.} \bibnamefont{Bronold}},
  \bibnamefont{and} \bibinfo{author}{\bibfnamefont{H.}~\bibnamefont{Fehske}},
  \bibinfo{journal}{Phys. Rev. B} \textbf{\bibinfo{volume}{85}},
  \bibinfo{pages}{075323} (\bibinfo{year}{2012}{\natexlab{a}}).

\bibitem[{\citenamefont{Heinisch
  et~al.}(2012{\natexlab{b}})\citenamefont{Heinisch, Bronold, and
  Fehske}}]{HBF12109}
\bibinfo{author}{\bibfnamefont{R.~L.} \bibnamefont{Heinisch}},
  \bibinfo{author}{\bibfnamefont{F.~X.} \bibnamefont{Bronold}},
  \bibnamefont{and} \bibinfo{author}{\bibfnamefont{H.}~\bibnamefont{Fehske}},
  \bibinfo{journal}{{Phys. Rev. Lett.}} \textbf{\bibinfo{volume}{109}},
  \bibinfo{pages}{243903} (\bibinfo{year}{2012}{\natexlab{b}}).

\bibitem[{\citenamefont{R\"opcke et~al.}(2006)\citenamefont{R\"opcke, Lombardi,
  Rousseau, and Davies}}]{RLR06}
\bibinfo{author}{\bibfnamefont{J.}~\bibnamefont{R\"opcke}},
  \bibinfo{author}{\bibfnamefont{J.}~\bibnamefont{Lombardi}},
  \bibinfo{author}{\bibfnamefont{G.}~\bibnamefont{Rousseau}}, \bibnamefont{and}
  \bibinfo{author}{\bibfnamefont{P.~B.} \bibnamefont{Davies}},
  \bibinfo{journal}{Plasma Sources Sci. Technol.}
  \textbf{\bibinfo{volume}{15}}, \bibinfo{pages}{S148} (\bibinfo{year}{2006}).

\bibitem[{\citenamefont{Carstensen et~al.}(2011)\citenamefont{Carstensen, Jung,
  Greiner, and Piel}}]{CJ11}
\bibinfo{author}{\bibfnamefont{J.}~\bibnamefont{Carstensen}},
  \bibinfo{author}{\bibfnamefont{H.}~\bibnamefont{Jung}},
  \bibinfo{author}{\bibfnamefont{F.}~\bibnamefont{Greiner}}, \bibnamefont{and}
  \bibinfo{author}{\bibfnamefont{A.}~\bibnamefont{Piel}},
  \bibinfo{journal}{Phys. Plasmas} \textbf{\bibinfo{volume}{18}},
  \bibinfo{pages}{033701} (\bibinfo{year}{2011}).

\bibitem[{\citenamefont{Fortov et~al.}(2004)\citenamefont{Fortov, Petrov,
  Usachev, and Zobnin}}]{FPU04}
\bibinfo{author}{\bibfnamefont{V.~E.} \bibnamefont{Fortov}},
  \bibinfo{author}{\bibfnamefont{O.~F.} \bibnamefont{Petrov}},
  \bibinfo{author}{\bibfnamefont{A.~D.} \bibnamefont{Usachev}},
  \bibnamefont{and} \bibinfo{author}{\bibfnamefont{A.~V.}
  \bibnamefont{Zobnin}}, \bibinfo{journal}{{Phys. Rev. E}}
  \textbf{\bibinfo{volume}{70}}, \bibinfo{pages}{046415}
  (\bibinfo{year}{2004}).

\bibitem[{\citenamefont{Ruppin}(1975)}]{Ruppin75}
\bibinfo{author}{\bibfnamefont{R.}~\bibnamefont{Ruppin}},
  \bibinfo{journal}{{Surface science}} \textbf{\bibinfo{volume}{73}},
  \bibinfo{pages}{140} (\bibinfo{year}{1975}).

\bibitem[{\citenamefont{Uberoi}(1980)}]{Uberoi80}
\bibinfo{author}{\bibfnamefont{C.}~\bibnamefont{Uberoi}},
  \bibinfo{journal}{Phys. Lett.} \textbf{\bibinfo{volume}{76A}},
  \bibinfo{pages}{69} (\bibinfo{year}{1980}).

\bibitem[{\citenamefont{Prodan et~al.}(2003)\citenamefont{Prodan, Radloff,
  Halas, and Nordlander}}]{Prodan03}
\bibinfo{author}{\bibfnamefont{E.}~\bibnamefont{Prodan}},
  \bibinfo{author}{\bibfnamefont{C.}~\bibnamefont{Radloff}},
  \bibinfo{author}{\bibfnamefont{N.~J.} \bibnamefont{Halas}}, \bibnamefont{and}
  \bibinfo{author}{\bibfnamefont{P.}~\bibnamefont{Nordlander}},
  \bibinfo{journal}{{Science}} \textbf{\bibinfo{volume}{302}},
  \bibinfo{pages}{419–422} (\bibinfo{year}{2003}).

\bibitem[{\citenamefont{Preston and Signorell}(2011)}]{Preston11}
\bibinfo{author}{\bibfnamefont{T.~C.} \bibnamefont{Preston}} \bibnamefont{and}
  \bibinfo{author}{\bibfnamefont{R.}~\bibnamefont{Signorell}},
  \bibinfo{journal}{{Proc. Nat. Acad. U. S. Sci}}
  \textbf{\bibinfo{volume}{108}}, \bibinfo{pages}{5532–5536}
  (\bibinfo{year}{2011}).

\bibitem[{\citenamefont{Mie}(1908)}]{Mie08}
\bibinfo{author}{\bibfnamefont{G.}~\bibnamefont{Mie}}, \bibinfo{journal}{{Ann.
  Phys. (Leipzig)}} \textbf{\bibinfo{volume}{25}}, \bibinfo{pages}{377}
  (\bibinfo{year}{1908}).

\bibitem[{\citenamefont{Stratton}(1941)}]{Stratton41}
\bibinfo{author}{\bibfnamefont{J.~A.} \bibnamefont{Stratton}},
  \emph{\bibinfo{title}{{Electromagnetic theory}}}
  (\bibinfo{publisher}{McGraw-Hill}, \bibinfo{year}{1941}).

\bibitem[{\citenamefont{Kerker}(1969)}]{Kerker69}
\bibinfo{author}{\bibfnamefont{M.}~\bibnamefont{Kerker}},
  \emph{\bibinfo{title}{{The scattering of light and other electromagnetic
  radiation}}} (\bibinfo{publisher}{Academic Press}, \bibinfo{year}{1969}).

\bibitem[{\citenamefont{Bohren and Huffman}(1983)}]{BH83}
\bibinfo{author}{\bibfnamefont{C.~F.} \bibnamefont{Bohren}} \bibnamefont{and}
  \bibinfo{author}{\bibfnamefont{D.~R.} \bibnamefont{Huffman}},
  \emph{\bibinfo{title}{{Absorption and Scattering of Light by small
  particles}}} (\bibinfo{publisher}{Wiley}, \bibinfo{year}{1983}).

\bibitem[{\citenamefont{Aden and Kerker}(1951)}]{AK51}
\bibinfo{author}{\bibfnamefont{A.~L.} \bibnamefont{Aden}} \bibnamefont{and}
  \bibinfo{author}{\bibfnamefont{M.}~\bibnamefont{Kerker}},
  \bibinfo{journal}{J. Appl. Phys.} \textbf{\bibinfo{volume}{22}},
  \bibinfo{pages}{1242} (\bibinfo{year}{1951}).

\bibitem[{\citenamefont{Wang et~al.}(2004)\citenamefont{Wang, Luk'yanchuk,
  Hong, Lin, and Chong}}]{WLH04}
\bibinfo{author}{\bibfnamefont{Z.~B.} \bibnamefont{Wang}},
  \bibinfo{author}{\bibfnamefont{B.~S.} \bibnamefont{Luk'yanchuk}},
  \bibinfo{author}{\bibfnamefont{M.~H.} \bibnamefont{Hong}},
  \bibinfo{author}{\bibfnamefont{Y.}~\bibnamefont{Lin}}, \bibnamefont{and}
  \bibinfo{author}{\bibfnamefont{T.~C.} \bibnamefont{Chong}},
  \bibinfo{journal}{{Phys. Rev. B}} \textbf{\bibinfo{volume}{70}},
  \bibinfo{pages}{035418} (\bibinfo{year}{2004}).

\bibitem[{\citenamefont{Schouten et~al.}(2004)\citenamefont{Schouten, Visser,
  and Lenstra}}]{SVL04}
\bibinfo{author}{\bibfnamefont{H.~F.} \bibnamefont{Schouten}},
  \bibinfo{author}{\bibfnamefont{T.}~\bibnamefont{Visser}}, \bibnamefont{and}
  \bibinfo{author}{\bibfnamefont{D.}~\bibnamefont{Lenstra}},
  \bibinfo{journal}{J. Opt. B} \textbf{\bibinfo{volume}{6}},
  \bibinfo{pages}{S404} (\bibinfo{year}{2004}).

\bibitem[{\citenamefont{Bashevoy et~al.}(2005)\citenamefont{Bashevoy, Fedotov,
  and Zheludev}}]{BFZ05}
\bibinfo{author}{\bibfnamefont{M.~V.} \bibnamefont{Bashevoy}},
  \bibinfo{author}{\bibfnamefont{V.~A.} \bibnamefont{Fedotov}},
  \bibnamefont{and} \bibinfo{author}{\bibfnamefont{N.~I.}
  \bibnamefont{Zheludev}}, \bibinfo{journal}{Opt. Express}
  \textbf{\bibinfo{volume}{13}}, \bibinfo{pages}{8372} (\bibinfo{year}{2005}).

\bibitem[{\citenamefont{Luk'yanchuk et~al.}(2006)\citenamefont{Luk'yanchuk,
  Tribelsky, and Ternovkij}}]{LT06}
\bibinfo{author}{\bibfnamefont{B.~S.} \bibnamefont{Luk'yanchuk}},
  \bibinfo{author}{\bibfnamefont{M.~I.} \bibnamefont{Tribelsky}},
  \bibnamefont{and} \bibinfo{author}{\bibfnamefont{V.~V.}
  \bibnamefont{Ternovkij}}, \bibinfo{journal}{J. Opt. Technol.}
  \textbf{\bibinfo{volume}{73}}, \bibinfo{pages}{060371}
  (\bibinfo{year}{2006}).

\bibitem[{\citenamefont{Luk'yanchuk et~al.}(2012)\citenamefont{Luk'yanchuk,
  Miroshnichenko, Tribelsky, Kivshar, and Khokhlov}}]{LMT12}
\bibinfo{author}{\bibfnamefont{B.~S.} \bibnamefont{Luk'yanchuk}},
  \bibinfo{author}{\bibfnamefont{A.~E.} \bibnamefont{Miroshnichenko}},
  \bibinfo{author}{\bibfnamefont{M.~I.} \bibnamefont{Tribelsky}},
  \bibinfo{author}{\bibfnamefont{Y.~S.} \bibnamefont{Kivshar}},
  \bibnamefont{and} \bibinfo{author}{\bibfnamefont{A.}~\bibnamefont{Khokhlov}},
  \bibinfo{journal}{{New J. Phys.}} \textbf{\bibinfo{volume}{14}},
  \bibinfo{pages}{093022} (\bibinfo{year}{2012}).

\bibitem[{\citenamefont{Landau and Lifschitz}(1984)}]{LL84}
\bibinfo{author}{\bibfnamefont{L.~D.} \bibnamefont{Landau}} \bibnamefont{and}
  \bibinfo{author}{\bibfnamefont{E.~M.} \bibnamefont{Lifschitz}},
  \emph{\bibinfo{title}{{Electrodynamics of Continuous Media}}}
  (\bibinfo{publisher}{Butterworth-Heinemann}, \bibinfo{year}{1984}).

\bibitem[{\citenamefont{Hofmeister et~al.}(2003)\citenamefont{Hofmeister,
  Keppel, and Speck}}]{Hofmeister03}
\bibinfo{author}{\bibfnamefont{A.~M.} \bibnamefont{Hofmeister}},
  \bibinfo{author}{\bibfnamefont{E.}~\bibnamefont{Keppel}}, \bibnamefont{and}
  \bibinfo{author}{\bibfnamefont{A.~K.} \bibnamefont{Speck}},
  \bibinfo{journal}{{Mon. Not. R. Astron. Soc.}}
  \textbf{\bibinfo{volume}{345}}, \bibinfo{pages}{16} (\bibinfo{year}{2003}).

\bibitem[{\citenamefont{Palik}(1985)}]{Palik85}
\bibinfo{author}{\bibfnamefont{E.~D.} \bibnamefont{Palik}},
  \emph{\bibinfo{title}{{Handbook of Optical Constants of Solids}}}
  (\bibinfo{publisher}{Academic}, \bibinfo{year}{1985}).

\bibitem[{\citenamefont{Barker}(1963)}]{Barker63}
\bibinfo{author}{\bibfnamefont{A.~S.} \bibnamefont{Barker}},
  \bibinfo{journal}{{Phys. Lett.}} \textbf{\bibinfo{volume}{132}},
  \bibinfo{pages}{1474} (\bibinfo{year}{1963}).

\end{thebibliography}

\end{document}